\documentclass{aa}
\usepackage[utf8]{inputenc}
\usepackage[T1]{fontenc}
\usepackage{amsmath,amssymb,amstext}
\usepackage{gensymb}
\graphicspath{{img/}}
\usepackage{xspace}
\usepackage{natbib}
\usepackage{xcolor}
\usepackage{pgfplotstable}
\usepackage{placeins}
\pgfplotstableset{
  col sep=comma,
  every head row/.style={before row={\hline},after row=\hline,},
  every last row/.style={after row=\hline},
}
\pgfplotsset{compat=1.16}
\makeatletter
\def\pgfmath@error#1#2{\vrule depth 1in}
\def\numorNaN#1{%
\setbox0\hbox{#1}%
\ifdim\dp0<1in\box0 \else\null\fi}
\def\bracketorNaN#1{%
\setbox0\hbox{#1}%
\ifdim\dp0<1in$^{+\footnotesize{\box0} }$\else\null\fi}
\usepackage{subcaption}

\usepackage{datatool} 

\usepackage[tableposition=top]{caption}
\usepackage{float}
\floatstyle{plaintop} 
\restylefloat{table}

\usepackage{multirow}

\newcommand{\eq}[1]{Eq.\,(\ref{#1})}
\newcommand{\eqs}[2]{Eqs.\,(\ref{#1}) and (\ref{#2})}
\newcommand{\fig}[1]{Fig.\,\ref{#1}}

\newcommand{\sect}[1]{Sect.\,\ref{#1}}

\newcommand{\app}[1]{Appendix\,\ref{#1}}
\newcommand{\tab}[1]{Table\,\ref{#1}}

\def\batman{\texttt{batman}\xspace}

\def\pytmo{Pytmosph3R\xspace}

\def\waspoto{WASP-121b\xspace}
\def\wasptn{WASP-39b\xspace}
\def\HJs{Hot Jupiters\xspace}
\def\HJ{Hot Jupiter\xspace}
\def\UHJs{Ultra Hot Jupiters\xspace}
\def\UHJ{Ultra Hot Jupiter\xspace}

\def\Rp{\ensuremath{R_P}}
\def\rprs2{\ensuremath{(R_P/R_S)^2}}
\def\Rs{\ensuremath{R_S}}
\def\Rsun{\ensuremath{R_\odot}\xspace}
\def\Ds{\ensuremath{r_S}}
\def\As{\ensuremath{\theta_S}}
\def\ra{\ensuremath{r_1}}
\def\rb{\ensuremath{r_2}}
\def\aa{\ensuremath{\theta_1}}
\def\ab{\ensuremath{\theta_2}}

\def\rc{\ensuremath{r}}

\def\phase{\ensuremath{\phi}}
\def\sma{\ensuremath{a}}

\def\nradial{\ensuremath{N_{r}}\xspace}
\def\ntheta{\ensuremath{N_{\theta}}\xspace}

\def\Nt{\ensuremath{N_\mathcal{T}}\xspace}
\def\Np{\ensuremath{N_{\phase}}\xspace}

\def\Negress{\ensuremath{N_{\textnormal{egress}}}\xspace}
\def\Ningress{\ensuremath{N_{\textnormal{ingress}}}\xspace}

\def\coordtrans{\Gamma_\mathcal{T}}

\def\cell{(\ra, \rb, \aa, \ab)}

\def\tr{\tau_{\rc,\theta}^{\lambda,\phase}}
\def\Tr{\ensuremath{\mathcal{T}}\xspace}
\def\trans{\ensuremath{\Tr_{\rc,\theta}^{\lambda,\phase}}\xspace}
\def\Sp{S_{\rc,\theta}^\phase}

\newcommand{\diff}{\mathop{}\!\mathrm{d}}

\newcommand{\balign}[1]{
\begin{align}
#1
\end{align}}

\usepackage{hyperref}
\hypersetup{colorlinks=true,citecolor=blue,linkcolor=blue!80!black,urlcolor=blue!80!black}
\begin{document}

\title{Signature of the atmospheric asymmetries of hot and ultra-hot Jupiters in lightcurves}
\titlerunning{Influence of asymmetries on lightcurves}

\author{Aur\'{e}lien Falco\inst{1,2,3}
\and J\'{e}r\'{e}my Leconte\inst{4}
\and Alexandre Mechineau\inst{4}
\and William Pluriel\inst{5}
}

\institute{
  Laboratoire de Météorologie Dynamique, IPSL, CNRS, Sorbonne Université, Ecole Normale Supérieure, Université PSL, Ecole Polytechnique, Institut Polytechnique de Paris, 75005 Paris, France\\
  \email{aurelien.falco@lmd.ipsl.fr}
  \and
  IPGP, Université Paris Cité, Université Paris-Saclay, CEA, CNRS, AIM, F-91191, Gif-sur-Yvette, France
  \and
  Laboratoire AIM, CEA, CNRS, Univ. Paris-Sud, UVSQ, Université Paris-Saclay, F-91191 Gif-sur-Yvette, France
  \and
  Laboratoire d'Astrophysique de Bordeaux, Univ. Bordeaux, CNRS, B18N, all\'{e}e Geoffroy Saint-Hilaire, 33615 Pessac, France
  \and
  Observatoire Astronomique de l’Universit\'{e} de Gen\`{e}ve, département d’astronomie Chemin Pegasi 51, CH-1290 Versoix, Switzerland 
}

\date{\today}

\abstract{With the new generation of space telescopes such as the James Webb Space Telescope (JWST), it is possible to better characterize the atmospheres of exoplanets.
The atmospheres of Hot and \UHJs are highly heterogeneous and asymmetrical. 
The difference between the temperatures on the day-side and the night-side is especially extreme in the case of \UHJs.
We introduce a new tool to compute synthetic lightcurves from 3D GCM simulations, developed in the \pytmo framework.
We show how rotation induces a variation of the flux during the transit that is a source of information on the chemical and thermal distribution of the atmosphere. 
We find that the day-night gradient linked to \UHJs has an effect close to the stellar limb-darkening, but opposite to tidal deformation. 
We confirm the impact of the atmospheric and chemical distribution on variations of the central transit time, though the variations found are smaller than that of available observational data, which could indicate that the east-west asymmetries are underestimated, due to the chemistry or clouds.
}

\maketitle

\section{Introduction}
\label{sec:introduction}

One-dimensional plane parallel geometry is sufficient to explain low resolution data in the case of cold planets, since the atmosphere is more or less homogeneous \citep{MacDonald2020}.
However, in the case of \HJs, this assertion does not hold anymore.
Such planets, tidally-locked and very close to their host star, present a strong dichotomy between the inflated hot day-side and the cold and shrunken night-side \citep{showmann2002,showman2008,menou2009,Wordsworth2011,heng2011,charnay2015,kataria2016, drummond2016, tan2019,pluriel2020strong,Pluriel2022}.
It is then necessary to use more-than-one dimensional models to explain spectral features.

In the case of transmission spectroscopy, the sampled atmospheric space is limited to a narrow region around the terminator \citep{brown2001, kreidberg2018}, and we may extract from the signal information from the day and the night-side \citet{fortney2010, caldas2019effects, Wardenier2022}.
Morning-evening asymmetries may also have an impact on the spectrum \citep{line2016,parmentier2016,MacDonald2020,espinoza2021}.
Even so, retrieval techniques, i.e., the inference of input parameters for the atmospheric simulation based on the fitting of a spectral observation, have mostly relied on 1D models until quite recently.
Studies have now started to include more dimensions \citep{MacDonald2021trident,Nixon2022aura3D,Zingales2022} for the fitting of observations.

Observations not only have spectral information, but also temporal information, i.e., the observed flux varies over time, and especially during the transit of one or more exoplanets.
\cite{koning2022impact} have also shown that stellar flares are also expected to have a time-dependent impact on transmission spectra (this effect will not be considered in this study).

What are the information that lightcurves (variations of the flux during a transit of an exoplanet) may provide?
Through time-resolved high-resolution spectroscopy, it is possible to detect species by cross-correlation of the features of each species \citep{prinoth2023time}.
\cite{prinoth2023time} emphasize that the observational data is expected to fluctuate during the transit and is influenced by the characteristics of the atmosphere such as the thermal dissociation occurring at very high temperatures on the day side of \UHJs.
Using low-resolution spectral information, lightcurves can provide evidence of morning and evening asymmetries
\citep{line2016,vonparis2016inferring}.
They also allow us to probe other regions of the atmospheres, due to the rotation of the planet.
Lightcurve models have historically relied on 1D spherical models \citep{kreidberg2015batman,tsiaras2016}, though recent tools extend the concept to different shapes and dimensions \citep{maxted2016ellc,akinsanmi2019,jones2020,Barros2022,grant2022transmission}. 

There is a similar trend in emission spectroscopy. For example \cite{chub2022exoplanet} rely on a 3D model to study phasecurves, enabling them to extract information such as a shift of the hot-spot in the simulation.

In this work, we consider 3D GCM simulations \citep{Wordsworth2011,Leconte2013}, from which we can extract information such as the temperature map and the chemical composition of the atmosphere.
\cite{caldas2018etude,caldas2019effects} introduced a tool, namely \pytmo, which produces transmission spectra from the simulation data.
A new implementation of \pytmo has later been introduced, more user-friendly and open-source, namely \pytmo~2.0 \citep{falco2022toward}.
The present paper is linked to a new release (i.e., 2.2) which introduces lightcurves, among other functionalities.
\sect{sec:forward_model} describes the transit model parameters and details.
\sect{sec:1D} shows the biases of a homogeneous spherical 1D lightcurve model (a solid sphere, or a completely opaque ball), as generally used by the literature, to a full 1D atmospheric model.
It is followed by a discussion (\sect{lc_retrievals}) on the impact of a 2D day-night temperature gradient approximation, as well as a full 3D GCM simulation on the shape of the resulting lightcurve, when taking into account rotation or when it is ignored.
This has been studied for a \HJ (\wasptn) and an \UHJ (\waspoto), both cases being presented in \sect{hot_ultra_hot}.
We compare the effect of the rotation of a tidally-locked planet to the stellar limb-darkening in \sect{atmosphere_vs_ld}, and to the ellipsoid approximation \citep{maxted2016ellc,Barros2022} in \sect{atmosphere_vs_ellipsoid}.

\section{Transit model}
\label{sec:forward_model}

\begin{figure}[ht]\centering
\includegraphics[width=.2\textwidth]{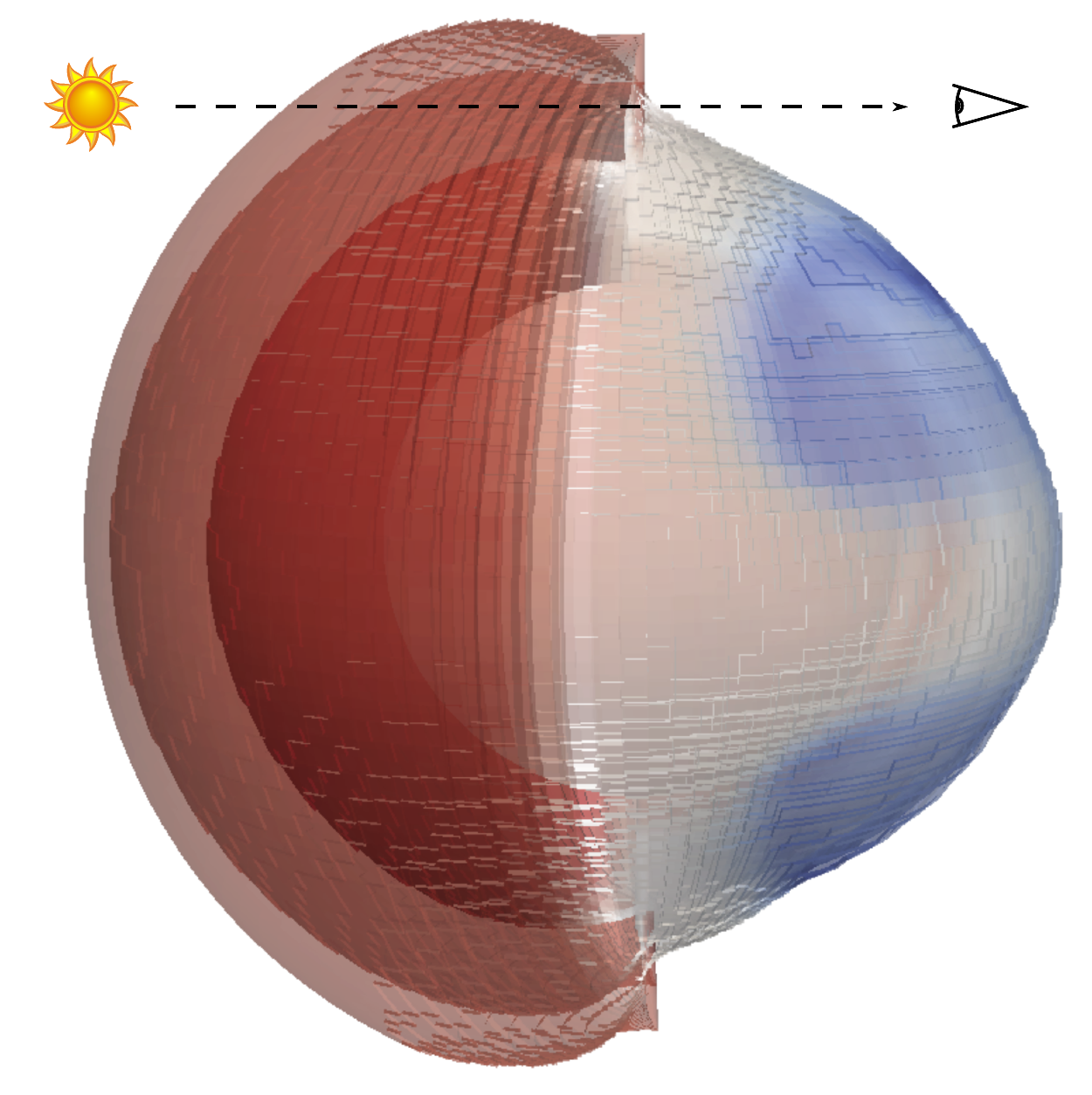}
\caption{Ray of light crossing the atmosphere of \waspoto. The color map indicates the local temperature, ranging from around 400~K (in blue) to 3300~K (in red). The inflation of the day-side is directly due to the increase of the scale height (due to the hydrostatic equilibrium). This simulation includes a super rotating equatorial jet (visible in gray).}
\label{fig:wasp121_rays}
\end{figure}

We briefly reintroduce here the method that we use to compute the transit depth of an exoplanet.
Let us consider a situation like the one represented in \fig{fig:wasp121_rays}.


$\coordtrans$ is defined as the polar coordinate system of the transmittance grid $\Tr$, located in the plane of the sky, orthogonal to the planet-observer axis.
The points in this coordinate system will be noted $(\rc, \theta)$.
In this coordinate system, the point ($\Rp$, 0) is the projection of the northenmost point on the planet of radius $\Rp$ on the transmittance grid.
The transmittance grid \Tr is composed of \nradial radial points and \ntheta angular points.

The rays of light crossing the atmosphere are partially absorbed by the interaction of the rays with the molecules present in the atmosphere.
The opacity of each ray may be calculated through the optical depth, which has already been detailed in \cite{caldas2019effects,falco2022toward}. 

\subsection{Transmittance map and spectrum}
\label{spectrum}

A transmittance map may be computed from the optical depth $\tau$ via
\begin{equation}
	\trans = e^{-\tr},
	\label{eq:transmittance}
\end{equation}
for each ray $(\rc, \theta)$ in the transmittance grid, wavelength $\lambda$ and phase $\phase$.
In the case of a homogeneous stellar disc, the relative dimming of the stellar flux of the planet and its atmosphere during transit is given by
\begin{equation}
	\Delta_\lambda(\phase) = \frac{{\pi R_p}^2 + \sum\limits_{\rc} \sum\limits_{\theta} \left(1 - e^{-\tr}\right) \Sp }{{\pi R_s}^2},
	\label{eq:integral}
 \end{equation}
with
$\Sp = 2 \pi (\rc + \frac{\diff\rc}{2})\diff\rc / \ntheta$, for phases $\phase$ when the planet is completely in front of the star.
$R_p$ is the radius of the planet, $R_s$ the radius of the star, $\diff\rc$ the distance between two consecutive $\rc$.

However, during the ingress and egress, the surface area of $\Sp$ will change according to its intersection with the stellar disc.
The details of this calculation is detailed in \app{sec:intersection}.

In this study, lightcurves have been generated for 30000 wavelength points between 0.5 and 10 $\mu m$.

\subsection{Transit parameters}
\label{sec:transit_model}

The transit is parameterized by the inclination of the orbit of the planet, its distance to the star $\sma$ (i.e., the semi-major axis, or the orbit radius since we consider the orbit to be spherical in the current study), and the number of phases at which we calculate the forward model.

The calculation of the transmittance map $\trans$ is quite costly, so we calculate $\Nt$ transmittances maps, a number that can be different from the number of phases $\Np$ for which we calculate \eq{eq:integral}.
In practice, we therefore calculate $\trans$ for phases
\begin{equation}
  \phase_\Tr = \{ \phase_j , \forall j \in [1, \Nt]   \},
  \label{eq:phases_transmittances}
\end{equation}
preferably with a low \Nt, and using these reference points $\phase_\Tr$, we interpolate $\trans$ for the phases $\phase_\lambda$:
\begin{equation}
  \phase_\lambda = \{ \phase_i , \forall i \in [1, \Np]   \},
  \label{eq:phases}
\end{equation}
which enables us to calculate \eq{eq:integral} for all phases $\phase_\lambda$.
The advantage of doing so is that \Np can be much larger than \Nt and we can therefore compute numerous timesteps without enduring the full cost of the computation of a transmittance map for each and every one of them.

An alternative method to improve speed without significantly compromising accuracy is to distribute the points $\phase_\lambda$ in a way that allocates more points for segments of the transit where the lightcurve is expected to exhibit a more intricate fluctuation.
Considering homogeneous stellar discs in this study, this will take place during the ingress and egress.
We consider the orbit to be spherical for now.
The phase at which the atmosphere will emerge from the transit (when it is not superimposed with the stellar disc anymore), i.e., the start of the egress, is equal to
\begin{equation}
  \phase_{\textnormal{start}} = \arcsin\left(\frac{\Rs-\Rp+z_{max}}{\sma}\right),
  \label{eq:start_egress}
\end{equation}
where $z_{max}$ is the maximum altitude of the simulation, and $\sma$ is the planet's orbit radius.
Similarly, the egress stage will stop at phase
\begin{equation}
  \phase_{\textnormal{stop}} = \arcsin\left(\frac{\Rs-\Rp-z_{max}}{\sma}\right).
  \label{eq:end_egress}
\end{equation}
These functions are valid only for circular orbits.
They also depend on the scale of the simulation and the lower pressures for which the atmosphere has been computed.

We consider the egress to take place between $\phase_{\textnormal{start}}$ and $\phase_{\textnormal{end}}$, the ingress between and $-\phase_{\textnormal{end}}$ and $-\phase_{\textnormal{start}}$.
The egress space defined by these points is discretized into $\Negress$ phases, the ingress into $\Ningress$ phases, and the rest of the phases ($\Np - \Negress - \Ningress$) discretize the plateau in-between, when the entire atmosphere of the planet is completely superimposed with the stellar disc.

In practice we will set $\Ningress = \Negress$, and the implementation sets $\Negress = \Np / 3$ by default.
Setting a higher phase resolution for the egress/ingress allows us to better replicate the fineness of the egress/ingress, during which the intersection between the atmosphere's projection and the stellar disc is going to change drastically at each phase.
During the rest of the transit, the main change in the transmittance is induced by the rotation of the planet. This can be approximated with a lower phase resolution \Np, interpolated between the $\Nt$ phases (for which a transmittance map has been computed).
The matter of choosing $\Nt$ is discussed in Appendix~\ref{sec:convergence}, considering a balance between accuracy and computational time.

\section{Homogeneously opaque solid sphere 1D fitting of the transmittance of a 1D atmosphere}
\label{sec:1D}

\begin{figure}[ht]\centering
\hfill
\includegraphics[width=.2\textwidth]{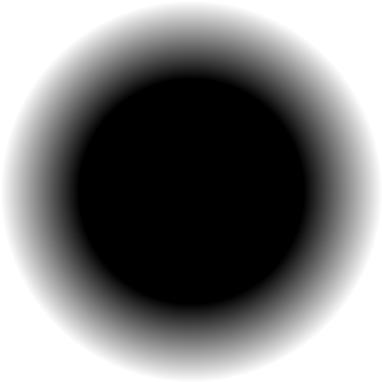}
\hfill
\includegraphics[width=.2\textwidth]{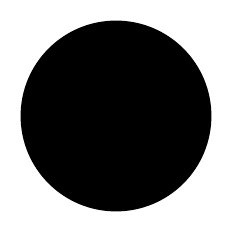}
\hfill
\caption{
Left: diffuse opacity due to the atmospheric content (\pytmo).
Right: homogeneously opaque solid sphere approximation (\batman).
}
\label{fig:diffuse_spherical_opacity}
\end{figure}

We study here the biases of one-dimensional and homogeneously spherical lightcurve models, meaning the entire solid sphere (or ball) is considered to be completely opaque.
We will show here that these models are able to retrieve the signal of a one-dimensional atmosphere almost consistently all along the transit, except during the ingress/egress.
To that end, we thus compare two models: 
\begin{enumerate}
    \item the 1D atmosphere, for which the lightcurve is generated via \pytmo~2.2 (see details in \tab{tab:1D_case});
    \item the homogeneously spherical model is generated via \batman.
\end{enumerate}
\batman is a python package introduced by \cite{kreidberg2015batman} for modeling 1D transit and eclipse lightcurves, which will be used as a reference point.
A simple representation of the two models is shown in \fig{fig:diffuse_spherical_opacity}.

\begin{table}[ht]\centering
  \begin{tabular}{|ccccc|}
  \multicolumn{5}{c}{1D}\\\hline
  &$T$ & [CO] & [H$_2$O] & Background \\\hline
  &2500 & $4 \times 10^{-4}$ & $10^{-4}$ & H$_2$ \\\hline\hline
  \multicolumn{5}{c}{2D}\\\hline
  $T_{day}$ & $T_{night}$ & $T_{deep}$ & $\beta$ & $P_{iso}$ \\\hline
  2800 & 750 & 2500 & 20\textdegree & $10^4$\\
  \hline
  \end{tabular}
  \caption{Parameters for the 1D and 2D atmospheres.}
  \label{tab:1D_case}
  \label{tab:2D_case}
\end{table}
\label{sec:static_2D}

\begin{figure}[ht]\centering
\includegraphics[width=.5\textwidth]{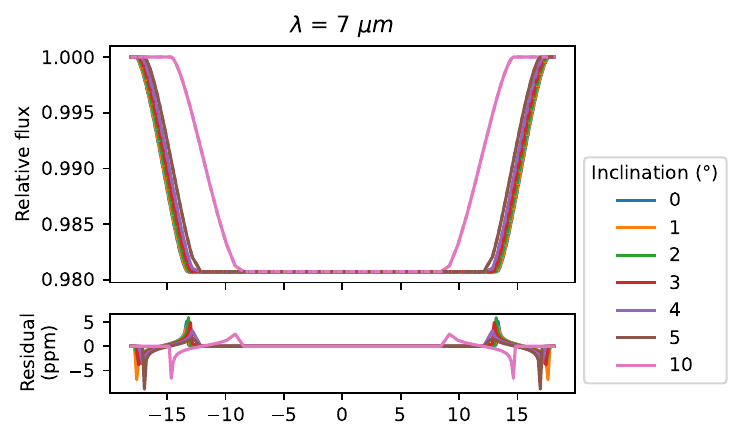}
\caption{Relative flux (\eq{eq:integral}) at a wavelength $\lambda = 7\mu m$ of the spherical approximation (\batman) compared to the diffuse opacity of a
1D (\tab{tab:1D_case}) atmospheric simulation (\pytmo).
See \fig{fig:diffuse_spherical_opacity} for the schematics.
The effect is similar in the case of a 2D day-night (\tab{tab:2D_case}) simulation (\pytmo) without rotation (i.e., $\Nt = 1$).
The same effect holds true for 3D cases, since the opacity of each case will decrease with the altitude, although it will be dominated by other effects.}
\label{fig:batman_vs_pytmo}
\end{figure}

The diffuse opacity of the atmosphere (which generally decreases with the altitude) calculated by \pytmo cannot be reproduced via a homogeneously opaque solid sphere.
This effect is only visible when the atmosphere partially absorbs the star light, i.e., during the ingress/egress.
This is shown in \fig{fig:batman_vs_pytmo}.

Though the integration of the opacity of the atmosphere over its apparent disc in \pytmo is equal to the circular surface calculated by \batman, the atmosphere calculated by \pytmo extends over a larger space than that of \batman, and is more transparent at high altitude, as shown by the schematics in \fig{fig:diffuse_spherical_opacity}.
Therefore, during the egress, the upper part of the atmosphere in \pytmo disappears from the transit before the equivalent disc in \batman does, and the opacity of \batman will thus be overestimated at the early stage of the egress.
Reversely, it is underestimated in the late stage of the egress.
The same holds true during the ingress.

The strength of this effect is relatively small (almost one order of magnitude) compared to the biases introduced by the spherical approximation in the presence of east-west or day-night asymmetries, which will be discussed in the following sections.

\section{From Hot to \UHJs}
\label{hot_ultra_hot}

In the later sections, we will study the effects of asymmetries in hot and \UHJs on the associated lightcurves.
For that purpose, we have selected three cases that we detail below.

\subsection{\UHJ \waspoto}

\begin{figure}[ht]\centering
\includegraphics[width=.5\textwidth]{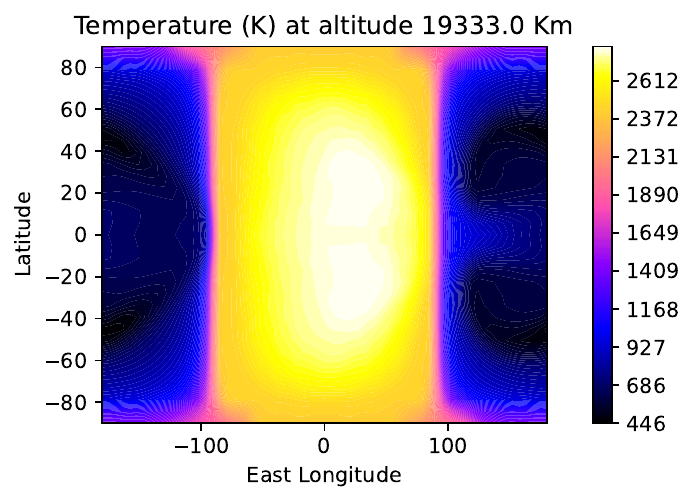}
\includegraphics[width=.46\textwidth]{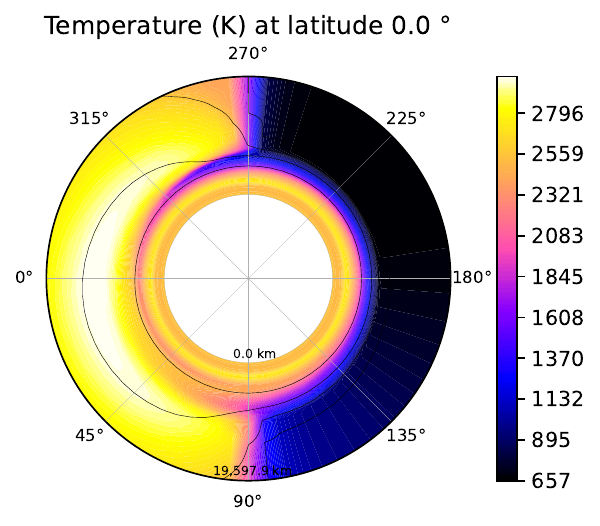}
\caption{Temperature maps of \waspoto. Top: high-altitudinal cut. Bottom: equatorial cut, for which the atmosphere scale has been multiplied by 10 for visual reasons.
In this simulation, the hot-spot is slightly shifted to the east.
}
\label{fig:wasp}
\end{figure}

\begin{figure}[ht]\centering
\includegraphics[width=.4\textwidth]{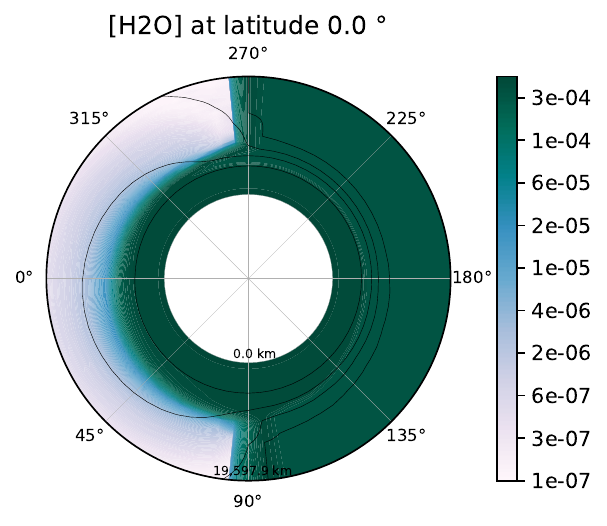}
\includegraphics[width=.4\textwidth]{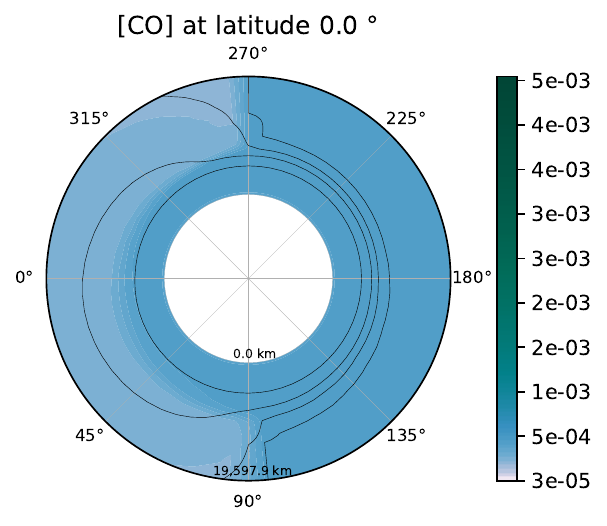}
\includegraphics[width=.4\textwidth]{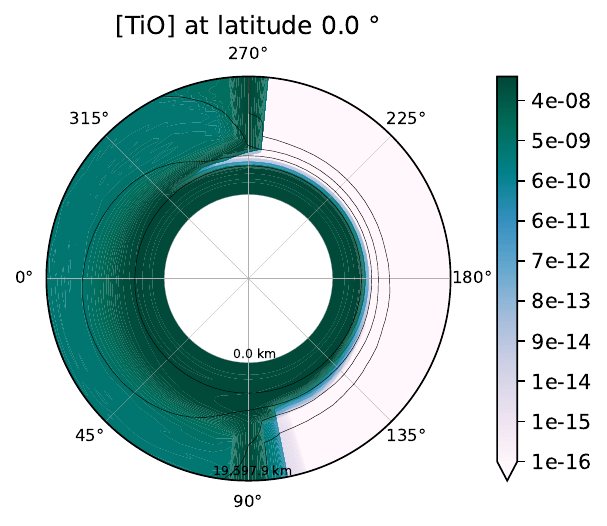}
\caption{Abundance maps of \waspoto. Top: H$_2$O. Middle: CO Bottom: TiO. All are equatorial cuts, for which the atmosphere scale has been multiplied by 10 for visual reasons.}
\label{fig:wasp_chemistry}
\end{figure}

In this study, we take a SPARC/MITgcm simulation \citep{showman2009} of \waspoto \citep{parmentier2018} as a reference for \UHJs.
Its temperature map is shown in \fig{fig:wasp}.
This simulation has a strong day-night temperature gradient and also a shift of the hot-spot towards the east.
The chemistry includes He, H,  H$_2$, H$_2$O, CO, TiO, VO, for which some of the relevant distribution maps are displayed in \fig{fig:wasp_chemistry}.

The star temperature is set to 6460~K, its radius to 1.458~\Rsun (solar radii) and the orbital radius of \waspoto to 0.025~au.

\subsection{2D day-night gradient}
\label{sec:rotation}
\label{sec:rotation_2D}

To be able to distinguish the signal due to day-night variations from the signal due to east-west variations in \UHJs, we will take as an example a 2D description of the atmosphere, for which the temperature is defined using \eq{eq:temperature2D} following a day-night gradient, as proposed by \cite{caldas2019effects}.
\setlength\arraycolsep{1pt} 
\begin{equation}
\label{eq:temperature2D}
\left\{ \begin{array}{lll}
	P > P_{iso}, & T = & T_{deep},\\
	\multirow{3}{*}{$ 
	P < P_{iso},
	\left\{ \begin{array}{l}
		2\alpha^* \geq \beta , \\
		2\alpha^* \leq -\beta, \\
		-\beta < 2\alpha^* < \beta,
	\end{array} \right. %
	$} 

	& T = & T_{day},\\
	& T = & T_{night},\\
	& T = & T_{night} \\
	& & + (T_{day} - T_{night}) 
	\frac{\alpha^*+\beta/2}{\beta},
\end{array} \right.
\end{equation}
The atmosphere is thus symmetrical around the star-observer axis, i.e., there is no morning-evening or east-west asymmetries.
The parameters used in this study for this 2D case are listed in \tab{tab:2D_case}.

\subsection{\HJ \wasptn}

\begin{figure}\centering
    \includegraphics[width=.5\textwidth]{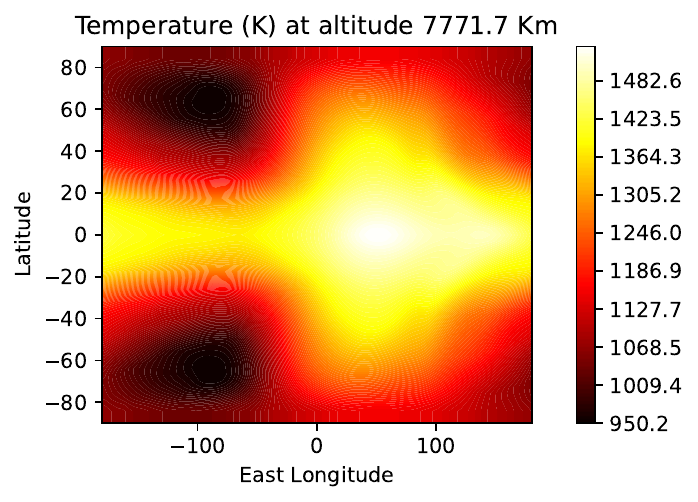}
    \includegraphics[width=.46\textwidth]{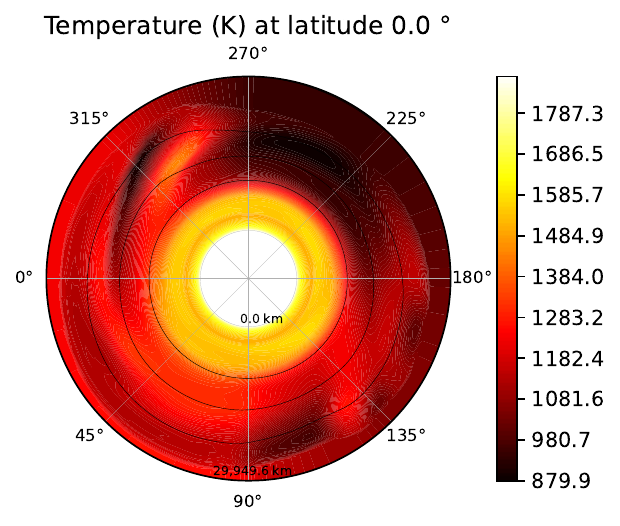}
    \caption{Temperature maps for \HJ \wasptn. Top: latitude/longitude map, at an altitude of 7772~km. Bottom: equatorial slice. The simulation goes up to an altitude of 29950~km.}
    \label{fig:t_maps_wasp_39b}
\end{figure}

\HJs and \UHJs have a notable difference concerning the strength of the equatorial jet and the resulting east-west asymmetries.
We will take here the example of the \HJ \wasptn, which is based on the Met Office Unified Model \citep{Drummond2018}.
For our simulation of \wasptn, we have used a constant chemistry and chosen two species that display a readily identifiable spectral signature, i.e., H$_2$O and CO$_2$, with a VMR of 3.57$\times10^{-3}$ for H$_2$O and 1.25$\times10^{-4}$ for CO$_2$, while we use a He/H$_2$ ratio of 0.2577.
These abundances are not extracted from JWST observational data of \wasptn.
Our C/O ratio (0.03) is quite low compared to the ratios (> 0.3) retrieved by \cite{rustamkulov2023early} for \wasptn and should thus be taken with caution. 
Note that there is no CO in our simulation, while it should be a substantial part of the carbon-bearing species and thus the C/O ratio.
This study is focused on the observational implications of thermal asymmetries of \HJs rather than simulating the exact case of \wasptn. 
Temperature maps of our simulation for \wasptn are displayed in \fig{fig:t_maps_wasp_39b}.
This figure shows the temperature structure of the atmosphere of a simulation corresponding to \wasptn, sliced at an altitude of 7800~km, and at the equatorial plane.
The simulation presents a strong jet going eastwards, with the hot-spot shifted by more than 60\textdegree. 
The temperature of the equatorial jet is also still very hot on the night-side (> 1300~K) at the altitude of 7800~km, and
the temperatures in the overall equatorial inner annulus (up to around $10^4$~km) are all higher than 1400~K.
Above this altitude, the cold-spot (900~K) is located between the west limb and the anti-stellar point (longitudes $\sim200$\textdegree-280\textdegree).
The west limb is more or less uniform, with a temperature around 1200~K.

This is quite different from the case of \waspoto, in which case the eastwards shift of the hot-spot is much less pronounced, and does not reach the night-side.
The day-night gradient is therefore much stronger.
The cold-spot also covers more angle on the night-side. 

The star temperature is set to 5326.6~K, its radius to 1~\Rsun and the orbital radius of \wasptn to 0.0486~au.

\section{Biases of 1D lightcurve retrievals of rotating Hot to \UHJs due to atmospheric asymmetries}
\label{lc_retrievals}

Using \pytmo~2.2, we can try to unravel the biases inherent to 1D lightcurve models by generating lightcurves of 2D or 3D atmospheric models.
For this purpose, we will again use the 1D solid-sphere model \batman \citep{kreidberg2015batman}, but also Juliet \citep{espinoza2019juliet}, a wrapper of multiple tools (including \batman), which offers the possibility of fitting data through Bayesian inference (naturally here we will focus on fitting lightcurves).

The parameters used for Juliet retrievals are listed in \tab{tab:parameters_juliet}.
\begin{table}[ht]\centering
\begin{tabular}{lp{4.2cm}c}\hline
Parameter & 
Description &
Prior 
\\ 
\hline\hline
$T_0 / P$ & 
Time of inferior conjunction (central transit time) normalized by the orbital period $P$ & 
$\mathcal{N}(0;0.01)$  
\\ 
$R_P/R_S$  & 
Planet radius normalized by the star radius &
$\mathcal{U}(0, 1)$  
\\ 
$\sma/R_S$  &
Orbit radius normalized by the star radius & 
$\mathcal{N}( 3.6,0.1)$    
\\
$q_1, q_2$ &
Limb darkening parameters (quadratic law) 
&
$\mathcal{U}(0, 1)$  
\\\hline
\end{tabular}
\caption{Parameters and their description for the retrievals with Juliet in \sect{lc_retrievals}
(all except last) and \sect{atmosphere_vs_ld}.}
\label{tab:parameters_juliet}
\tablefoot{\\
$P = 1.27492$ days,
$R_S = 1.458~R_\odot$,
$\mathcal{U}(a;b)$ is a uniform distribution between $a$ and $b$;  $\mathcal{N}(a;b)$ is a
normal distribution with mean $a$ and standard deviation $b$. }
\end{table}
They include the time of inferior conjunction (mid-transit), or central transit time, normalized by the orbital period of the planet and, the planet radius and the orbit radius (considering a circular orbit) normalized by the host star radius.

Appendix~\ref{phase_resolution} discusses the impact of the time or phase resolution in the input lightcurve calculated by \pytmo during a retrieval.
The parameters used by \pytmo for the calculation of the lightcurves are listed in \tab{tab_lc_params} (i.e, the ingress/egress are discretized with a higher resolution).
As explained in Appendix~\ref{phase_resolution}, all lightcurves have been re-interpolated over $\Np = 300$ phases before feeding it to the retrieval to be statistically more adequate.

\subsection{Biases during the ingress/egress for a translating \UHJ with a 3D atmosphere (no rotation)}
\label{sec:static_3D}

Let us start with a simple case, and assume that we can neglect rotation during the transit due to its brevity (we set the number of computed transmittances \Nt to 1).
We will use the term ``translation'' to refer to a transit for which rotation is not considered.
We take the case of the \UHJ \waspoto as an example and generate its lightcurve using \pytmo~2.2.
We also generate the corresponding 1D solid sphere, homogeneously opaque, setting the input planet radius of \batman to the apparent radius at mid-transit of the lightcurve of \waspoto .

\begin{figure}[ht]\centering
    \includegraphics[width=.5\textwidth]{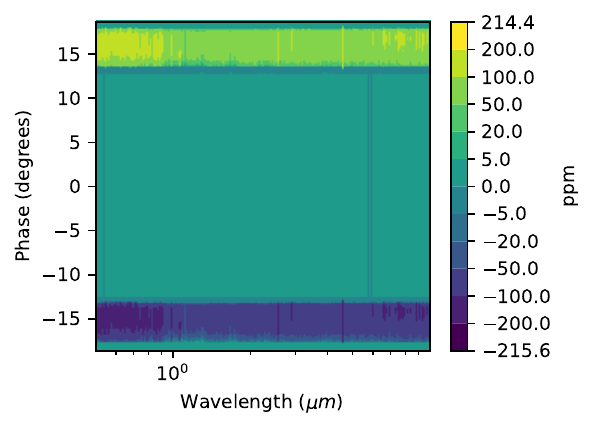}
    \caption{Difference between the flux (\eq{eq:integral}) of a 1D solid sphere  (\batman) and the flux of a 3D simulation of an \UHJ (\pytmo) without rotation (i.e., $\Nt = 1$). 
    The \UHJ is \waspoto (\fig{fig:wasp}).
    X-axis shows the wavelength; Y-axis shows the transit phase.
    }
    \label{fig:diff_batman_nt_1}
\end{figure}

The difference between the 1D (\batman) and 3D (\pytmo) lightcurves is shown in \fig{fig:diff_batman_nt_1}.
We can see on this figure the east-west asymmetry of the simulation.
The spherical model (\batman) finds an apparent disc larger than \pytmo during the ingress and overestimates it during the egress.
This effect is similar to Fig.~12 from \cite{vonparis2016inferring} for example.
This is due to the lower temperature in the west limb of the simulation, and higher temperature in the east limb as the hot-spot in the day-side is slightly shifted towards the east longitudes, as shown in \fig{fig:wasp}.
We will see that this effect can be partially reduced by changing the central time T$_0$ (\sect{central_time_variations}).

\subsection{The translation approximation}

\begin{figure} [ht]\centering
    \includegraphics[width=.45\textwidth]{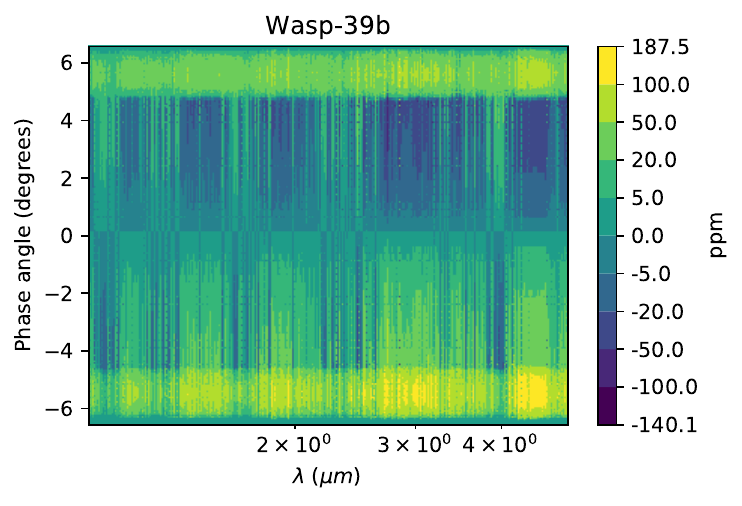}
    \includegraphics[width=.45\textwidth]{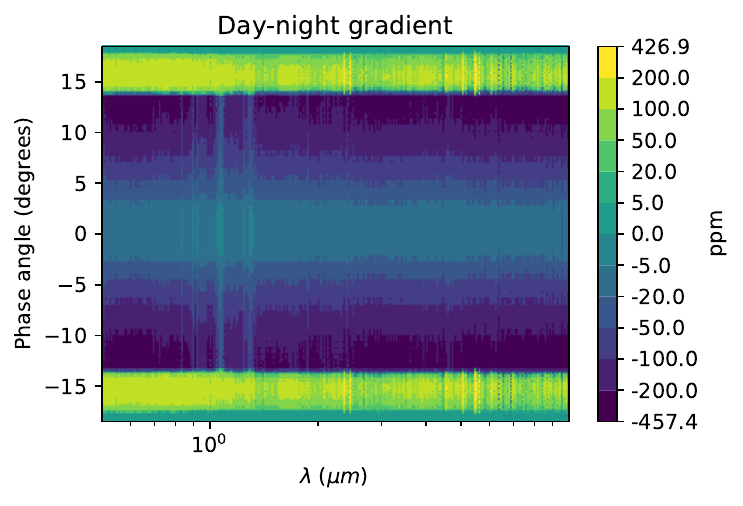}
    \includegraphics[width=.45\textwidth]{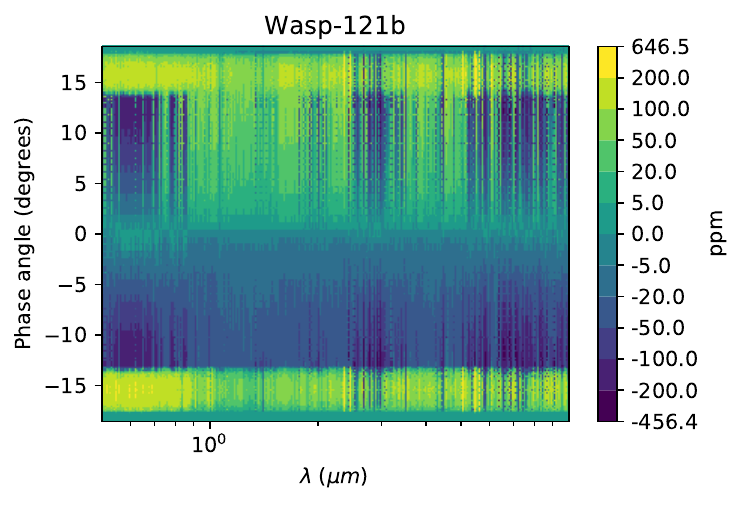}
    \caption{Residuals between translation ($\Nt = 1$) and rotation ($\Nt = 8$) fluxes.
    Top: \wasptn (\fig{fig:t_maps_wasp_39b}). 
    Middle: 2D day-night gradient (\tab{tab:2D_case}),  the east-west symmetry leads to the signal being identical at phases $\phase$ and $-\phase$.
    Bottom: \waspoto (\fig{fig:wasp}).
    X-axis shows the wavelength; Y-axis shows the transit phase.
    }
    \label{fig:residual_lc_wasp39b}
    \label{fig:diff_2D}
    \label{fig:diff_3D}
\end{figure}

Rotation is usually excluded from the current models, the planets being thought to be too cold and the transit too short for the rotation to usually have an effect on the lightcurve.
We will try to characterize here the effects of tidally-locked rotation that can be expected, from a pure geometrical perspective (i.e., ignoring Doppler effects).
In the ``translation'' case, the planet slides along the plane of the sky, ignoring the tidally-locked rotation.
\fig{fig:residual_lc_wasp39b} shows the difference between lightcurves generated for \wasptn, the 2D day-night gradient and \waspoto, each with and without rotation ($\Nt = 8$, and $\Nt = 1$, respectively).

In the case of \wasptn (top plot),
we can see that the signal is relatively stronger in the early stage rather than the late stage.
This is due to the fact that the cold-spot is located near the west longitudes ($\sim200$\textdegree-280\textdegree, see \fig{fig:t_maps_wasp_39b}).
This cold-spot moves towards the edge of the west limb during the late stage, accentuating the decrease in apparent radius in the west.
Indeed the scale-height of the atmosphere is linearly dependent on the temperature 
(H~=~RT/$mg$, with R standing for the ideal gas constant, $m$ the mean molar mass, $g$ the surface gravity and T the atmosphere temperature).
The cold-spot moves towards the anti-stellar point during the early stage, and is therefore less probed by stellar rays crossing the atmosphere in the star-observer axis. At the same time, the higher temperatures of the day-side move to the west limb, increasing its scale-height.
The effect is stronger for wavelengths with a high apparent radius (see \fig{fig_rprs_retrieved_wasp_39} or \fig{fig_rprs_retrieved_wasp_121} for example).
The signal during the ingress and egress are also boosted, reaching a peak of almost 200~ppm.
This peak during the ingress/egress is due to the overall day-night temperature difference.

\begin{figure*}[ht]\centering
\includegraphics[width=.31\textwidth]{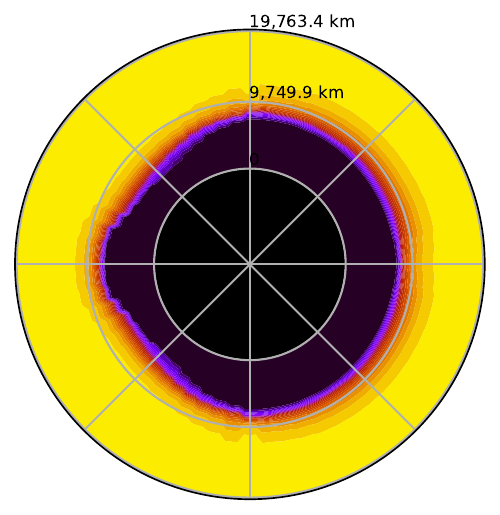}
\hfill
\includegraphics[width=.31\textwidth]{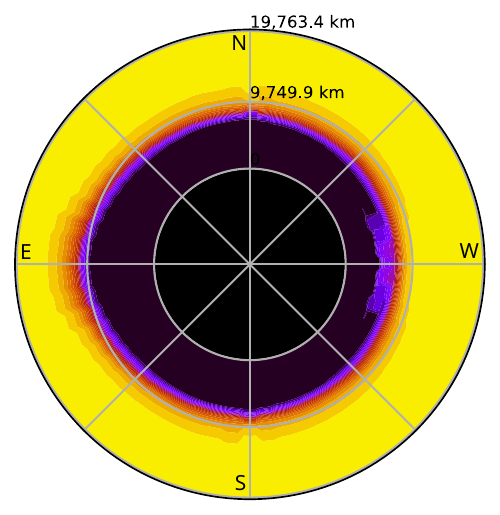}
\hfill
\includegraphics[width=.31\textwidth]{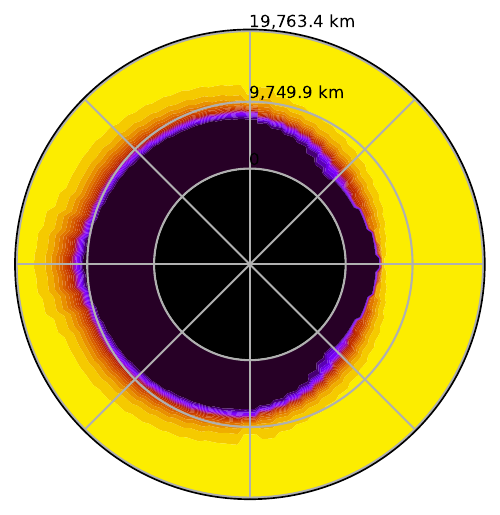}
\includegraphics[width=.05\textwidth]{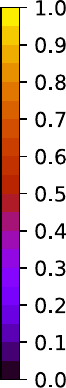}
\caption{Transmittance maps for phases -13, 0 and 13\textdegree, i.e., early, mid and late transit, respectively, at $\lambda$ = 0.6~$\mu m$ of \waspoto.
The atmosphere scale has been multiplied by 10 for visual reasons.
The inner black circle represents the planet core.
The total apparent surface is greater at mid-transit due to the hot day-side being apparent all around the limb. The night side is more apparent during the early and late stages (lower scale-height).
}
\label{fig_transmittances}
\end{figure*}

To understand this effect, let us focus on the 2D day-night gradient simulation (middle plot of \fig{fig:diff_2D}), for which the gradient between the day temperatures and the night temperatures is much stronger.
The lightcurves in this case are symmetric with respect to the mid-transit, meaning that $\Delta_\lambda(\phase) = \Delta_\lambda(-\phase)$.
The variations between different wavelengths are solely linked to the height of the absorption lines. 
In this case, the greatest differences between the rotation case and the translation case occur (by far) during the ingress and egress, due to the (hotter) day-side partially obscuring the star when the (colder) night-side is out of the transit zone.
For a visualization, see how the east limb is much larger (due to the day-side being more visible) in the last transmittance map (late transit) in \fig{fig_transmittances}.
This figure shows the example of \waspoto, although the same effect is happening for the 2D day-night gradient.
The west limb (which corresponds to a larger fraction of the night-side) is much smaller and is going to disappear first from the signal during the egress.
If we go back to \fig{fig:diff_2D}, we can also see on the 2D case that, in the early and late stages, the differences between rotation and translation can go as low as -450~ppm due to the projection average area of the atmosphere being smaller (see again the same transmittance maps in \fig{fig_transmittances}).
This difference follows a U-shape curve during the transit, excluding the ingress/egress (the difference is close to 0 at mid-transit but increasingly negative for phases far from mid-transit).

Our simulation of \waspoto presents a very strong day-night asymmetry, but also a smaller east-west asymmetry.
The U-shape signal of the day-night gradient (as discussed above for the 2D case) is again visible in the background in the 3D-case case (bottom plot of \fig{fig:diff_3D}), but the east-west asymmetry breaks the symmetry of the pattern.
The differences during the late-transit are positive.
They are due to the shift of the hot-spot on the day-side to the east, which moves towards the east limb during the late stage and is therefore more visible.
The differences amount for more than 100~ppm during the ingress/egress (reaching a peak of 600~ppm in the wavelengths with the larger apparent radii).
During the early/late stage, they can go as low as -450~ppm due to the day-night gradient and reach +100~ppm due to the shift of the hot spot.

Appendix~\ref{sec:convergence} also shows the errors and the times expected for different numbers of transmittances maps \Nt calculated during the transit (for faster calculation, interpolating between each phase, see \sect{sec:transit_model}), in the case of an \UHJ.

\subsection{Central time variations / east-west effects}
\label{central_time_variations}

We are interested here in studying the relationship between the atmosphere characteristics and the variations in the central time T$_0$ retrieved by the 1D spherical model of Juliet.

\begin{figure}
    \centering
    \includegraphics[width=.45\textwidth]{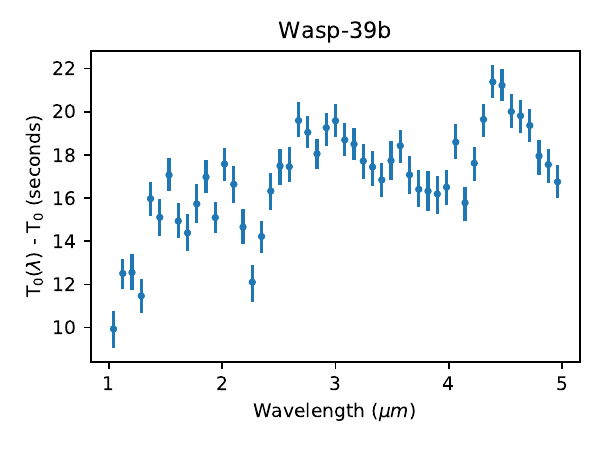}
    \includegraphics[width=.45\textwidth]{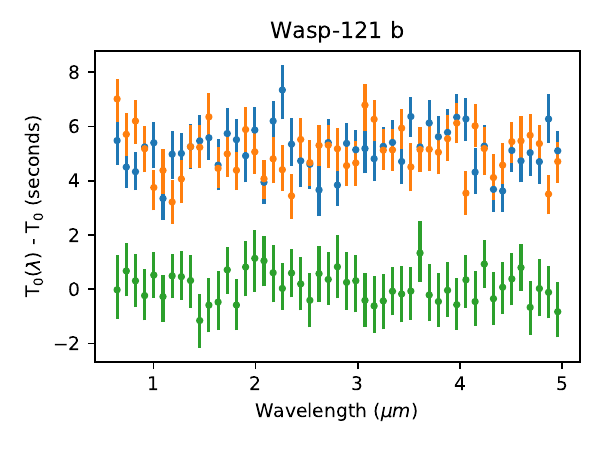}
    \includegraphics[width=.23\textwidth]{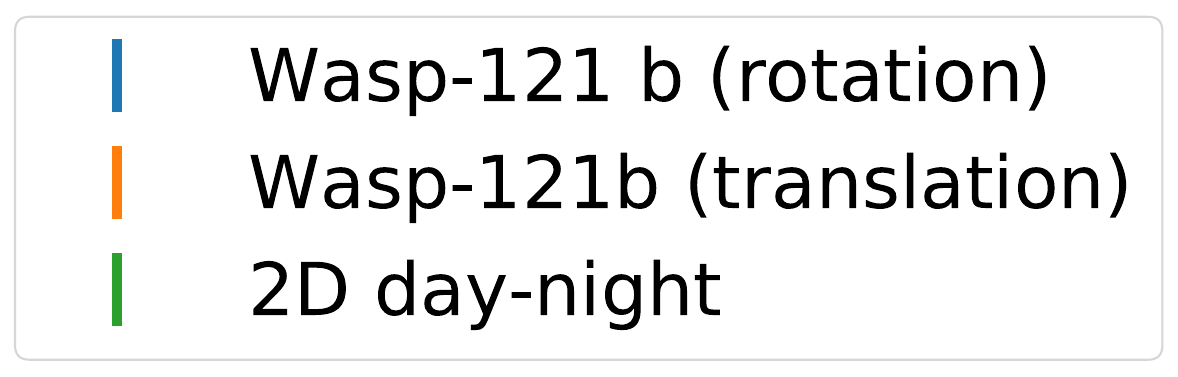}
    \caption{Time of inferior conjunction T$_0$ retrieved by Juliet for the tidally-locked \HJ \wasptn (top) and the \UHJ \waspoto (bottom, including legend) for wavelengths from 1 to 5 $\mu m$, at a resolution similar to NIRSpec.
    In the case of \waspoto, three cases are displayed (indicated by the colors): 
    1) 
    the lightcurve of the 3D simulation of \waspoto
    without rotation (i.e, translation)
    in orange, 
    2)
    the same 
    including rotation 
    in blue,
    3) a 2D day-night temperature gradient (symmetric transit) in green.
    }
    \label{fig_time_retrieved}
\end{figure}

\cite{rustamkulov2023early} have published the variations of the central transit time, or time of inferior conjunction, T$_0$ (see their Extended Data Fig.~3) for \wasptn. 
They show variations of T$_0$ between -50 and +60 seconds, depending on the wavelength (going from 0.5 to 6~$\mu m$).
We have generated a synthetic lightcurve with \pytmo for which we retrieve T$_0$, for wavelengths going from 1 to 5 $\mu m$.
We have done so for the \HJ \wasptn, considering a tidally-locked rotation, and for the \UHJ \waspoto.
The results are shown in \fig{fig_time_retrieved}.

In the case of the 2D day-night model, as shown in the middle plot of \fig{fig:diff_2D}, the shape of the transit is symmetrical before and after the mid-transit, i.e., $\Delta_\lambda(\phase) = \Delta_\lambda(-\phase)$, due to the symmetry of the simulation along the star-observer axis.
Therefore, retrievals by Juliet find no variation in the time of inferior conjunction T$_0$.

\fig{fig_time_retrieved} shows that the variations for the central time T$_0$ are stronger in the case of the \HJ \wasptn than in the case of the \UHJ \waspoto.
This is due to the fact that east-west asymmetries are stronger in the case of the \HJ than for the \UHJ.
The amplitude of the variations is around 5~seconds for all wavelengths (from 3 to 8 seconds) in the case of the \UHJ \waspoto, meaning that the retrieval method finds a solid-sphere equivalent planet that transits 5~seconds late compared to the actual planet.
It is 10~seconds at minimum  (around 1~$\mu m$) and going up to 22~seconds (around 4.5~$\mu m$) for the \HJ \wasptn.
For \wasptn, the spectral features are clearly visible. 
As for errors on the flux discussed above, wavelengths with larger apparent radii are also linked to a larger bias in the transit timing.

For the case of \wasptn, \cite{rustamkulov2023early} show an amplitude of more than 50~seconds on the same spectral range. 
Thus we manifestly do not account for the whole amplitude of the variations of T$_0$.
This could be due to cloud coverage or the chemical structure of the simulation.
As mentioned before, we have considered a quite simple constant chemistry for this simulation.
The observational \citep{rustamkulov2023early} and simulated data (this study) central time variations both follow the spectral features and especially the H$_2$O and CO$_2$ bands. 
Both data can also be fitted by a non-zero slope linear regression.
Let us assume a linear fit under the form $ax +b$, where $x$ is within $1-5~\mu m$.
In the case of the observational data \citep{rustamkulov2023early}, we obtain $a=2.07$ and $b=-4.77$ while we obtain $a=1.40$ and $b=12.68$ for our simulation of \wasptn.
The differences between the slope values ($a$) may be due to the stellar spectrum (if the real stellar temperature is different from ours, which is 5326.6~K) but also to our choice of chemistry, which do not correspond to the actual observations of \wasptn.
Indeed, a potential discrepancy in our simulation is the absence of CO, as there is a  CO band around 5~$\mu m$. The differences might also arise from an overestimated amount of H$_2$O, as the water bands dominate the wavelengths shortward of $3~\mu m$.

In our translation simulation of \waspoto (displayed in orange in the bottom plot of \fig{fig_time_retrieved})
the shift in the transit timing (of the order of 5~seconds for all wavelengths) is in line with previous results.
Indeed, \cite{line2016} found that lightcurve residuals could be reduced by a factor of 10 when shifting the transit timing of $\sim 10^{-5}$ and $\sim 10^{-4}$ days following the scale height difference between the morning and evening limbs of the planet.

The value found for T$_0$ in this translating case is also impacted by the chemistry.
For example, the value found for $\lambda = 0.6~\mu m$ in the translation case is higher than the other, due to the distribution of TiO in the simulation.
TiO is less symmetrically distributed than its counterparts H$_2$O and CO, and therefore its east-west asymmetry leads to an overestimation of T$_0$.
Other spectral features have less impact over the variations of T$_0$ and one might have difficulties in distinguishing the water or CO bands from the error bars.

When we account for the tidally-locked rotation of \waspoto (blue curve in \fig{fig_time_retrieved}), the central time is shifted by the same order of magnitude, around 5~seconds.
However, there are some differences, including in the spectral region dominated by the opacity of TiO (<~1$\mu$m), where the shift is slightly smaller.
This could be linked to the distribution of TiO in the atmosphere, which is indeed present at higher altitude on the east limb, but also in both terminators.
The fact that the planet rotates seems to reduce the difference between the east and west limbs, due to the fact that the oversized east limb (see \fig{fig:wasp_chemistry}) is reduced by being partly masked from the observer during the early and late stages, and the east-west differences are therefore less visible, compared to the translation case.
In the translation case, the east-west asymmetries present in both limbs stay constant throughout the whole transit.
Concerning the impact of the chemistry, we can hardly see the spectral signatures in the rotating case as well as in the translating case.
Spectral signatures are much more significant in the variations of T$_0$ for the \HJ \wasptn than for the \UHJ \waspoto.

\subsection{Reduced radius retrieved due to the rotation of the planet}
\label{sec_reduced_radius}

Due to the shape of the signal during the transit (U shape due to the day-night gradient, or slope due to east-west asymmetries), which cannot be simulated by a 1D spherical model, 1D retrievals will tend to average the observed radius over the whole duration of the transit.

\begin{figure} [h]\centering
    \includegraphics[width=.45\textwidth]{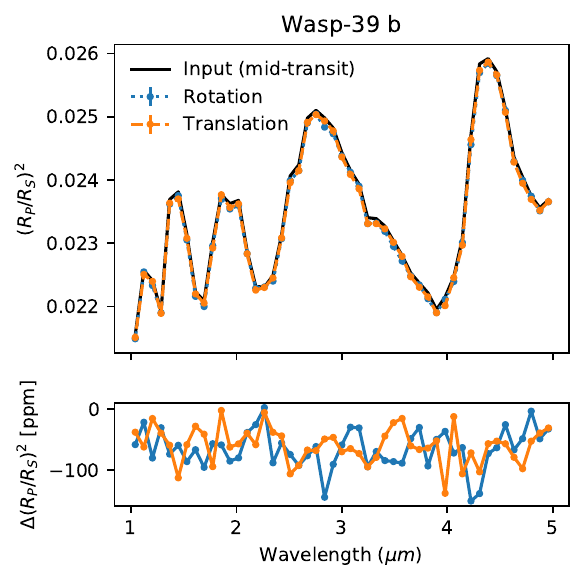}
    \caption{Retrieved values for \rprs2 by Juliet of \wasptn for the translation case (in orange, the planet does not rotate during the transit), the rotation case (in blue), and the input value of \rprs2 at mid-transit (in black).
    Error bars are included (less than 1~ppm).
    }
    \label{fig_rprs_retrieved_wasp_39}
\end{figure}

\begin{figure} [h]\centering
    \includegraphics[width=.45\textwidth]{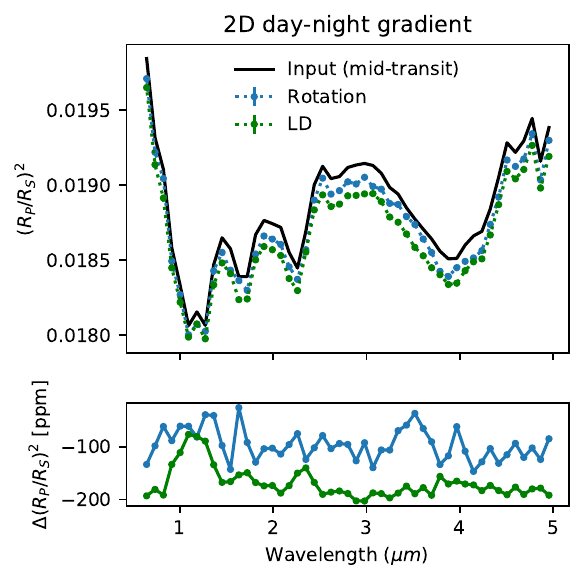}
    \caption{Retrieved values for \rprs2 by Juliet of the 2D day-night gradient for the rotation case (see \sect{sec_reduced_radius}), the same when also retrieving limb-darkening coefficients ("LD", see \sect{atmosphere_vs_ld}), and the input value of \rprs2 at mid-transit (in black).
    }
    \label{fig_rprs_retrieved_wasp_2D}
\end{figure}

\begin{figure} [h]\centering
    \includegraphics[width=.45\textwidth]{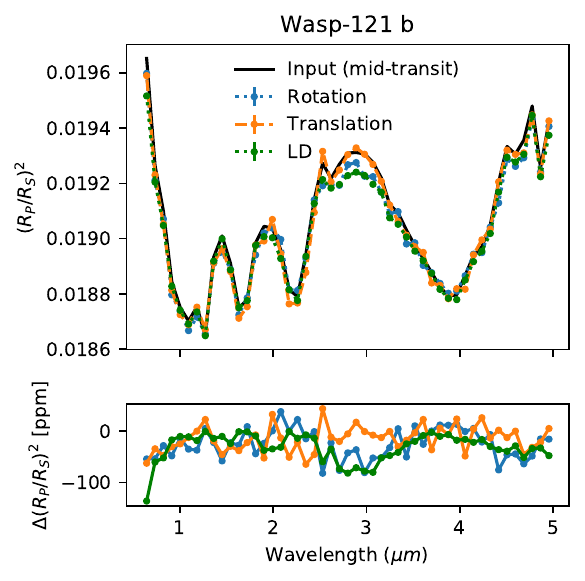}
    \caption{Same as \fig{fig_rprs_retrieved_wasp_2D} for \waspoto, with the retrievals of the rotation and translation cases (discussed in \sect{sec_reduced_radius}) and the retrieval using limb-darkening (discussed in \sect{atmosphere_vs_ld}).}
    \label{fig_rprs_retrieved_wasp_121}
\end{figure}

For the case of \wasptn, the radius retrieved for the rotation and the translation case, is displayed in \fig{fig_rprs_retrieved_wasp_39}.
For the 2D day-night case, the radius retrieved for the rotation and the translation case, is displayed in \fig{fig_rprs_retrieved_wasp_2D}.
For the case of \waspoto, the radius retrieved for the rotation and the translation case, is displayed in \fig{fig_rprs_retrieved_wasp_121}.

\begin{figure}[ht]\centering
\includegraphics[width=.49\textwidth]{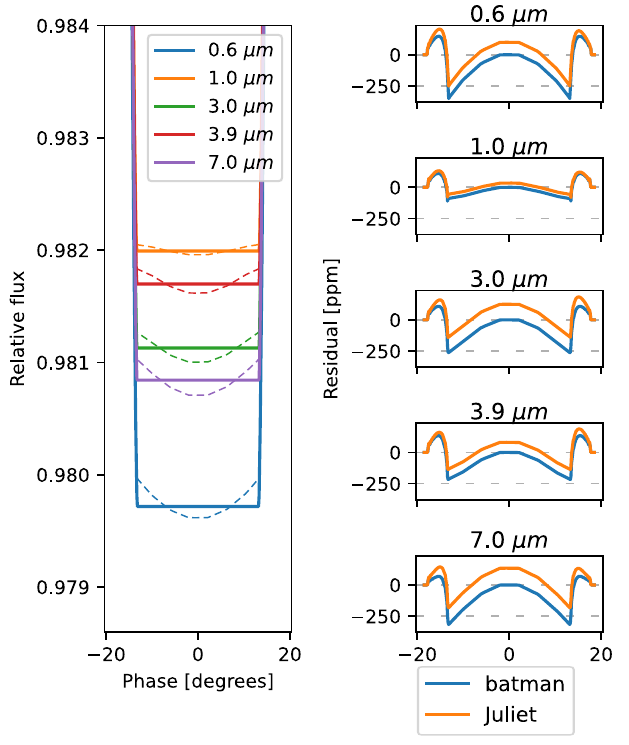}
\caption{2D rotation case. Left: Lightcurves generated by \pytmo (solid line) compared to
the fit provided by Juliet (dashed line, with retrieved values for $T_0$ and $R_P$).
Right: Residuals of Juliet (orange) to \pytmo, with \batman (assuming $T_0 = 0$ and $R_P$ taken at mid-transit from \pytmo) given as a reference.
Note that Juliet itself is using \batman as a forward model, the difference only comes from the retrieved values of $T_0$ and $R_P$.
}
\label{fig:juliet_fit_2D_rotate}
\end{figure}

One may notice that the retrieved values for the planet radius are almost always inferior to the input value.
This is due to the fact that the input value shown in the plot is taken at mid-transit. 
For lightcurves that have a U-shape, the average value of the signal will always be lower than the mid-transit value, since it is the maximum value of the whole transit.
This is more readily visible in \fig{fig:juliet_fit_2D_rotate}.
The input radius given to \batman is the maximum apparent radius $\Rp$ of the lightcurve generated via \pytmo, hence the residual of \batman is always equal to zero when the lightcurve of \pytmo reaches its minimum.
In this case, there is no meaningful value for $\Rp$ for neither \batman nor Juliet since the 1D model used cannot reproduce the non-linear curve shape during transit due to the asymmetries in the input model.

The differences between the Juliet retrieved lightcurves and the 2D day-night simulation are at almost constantly at their maximum at mid-transit (\fig{fig:juliet_fit_2D_rotate}), thus the retrieved apparent radii in \fig{fig_rprs_retrieved_wasp_2D} also have the largest errors among the three examples.
Indeed, in the cases of \wasptn and \waspoto, the lightcurves in some wavelengths follow a slope instead of a U-shape curve, and the maximum difference between the retrieved value of $R_P$ and its initial value is therefore not at mid-transit, but during the early and late stages (the retrieved value being an average of the overall slope, it is closer to the intermediate value of $R_P$ at mid-transit).

If we take a closer look at \fig{fig:juliet_fit_2D_rotate},
we can see that Juliet always overestimates the apparent radius during the early and late stages of the transit (which is equivalent to say that it underestimates the relative flux).
This is due to the fact that the projection of our 3D simulation of \waspoto over the observer's field of view is larger at mid-transit, as explained in previous sections (the hot temperatures from the day-side are contributing to the absorption all around the terminator), while being smaller during early and late stages, due to the hot day-side temperatures being replaced by the night-side's on the east and the west limb, respectively. 
See \fig{fig_transmittances} for a visualization.

During the ingress and egress, Juliet underestimates the apparent radius compared to the input simulation due to the day-side being the major part of the atmosphere obscuring the star (the night-side exits the transit before the day-side, and enters after the day-side).

In conclusion, a 1D model would be biased up to a few hundred of ppm depending on the current phase of the transit due to the rotation of the tidally-locked planet. 
One might need to take this effect into account in the calculation of the limb darkening associated to the star.

\section{Rotation of the atmospheric signal vs limb-darkening}
\label{atmosphere_vs_ld}

Since the 2D day-night gradient introduces a reduction of the signal during the early and late stages, the effect could be mistaken with the limb-darkening of the star.
When considering limb-darkening, the flux emitted by the star when the planet is obscuring the edge of the star is lower than when it obscurs the center.
Limb-darkening therefore also induces a U-shaped signal in the transit lightcurve, which is coincidentally similar to the effect of the day-night gradient of the planet, although the U-shape signal does not have the same properties.
We will consider here a quadratic law for the limb-darkening.

In this section, we retrieve the lightcurves generated via \pytmo for the 2D day-night gradient case and \waspoto.
The limb-darkening-specific parameters retrieved by Juliet are listed in 
\tab{tab:parameters_juliet}.

\begin{figure*} [ht]\centering
    \includegraphics[width=.48\textwidth]{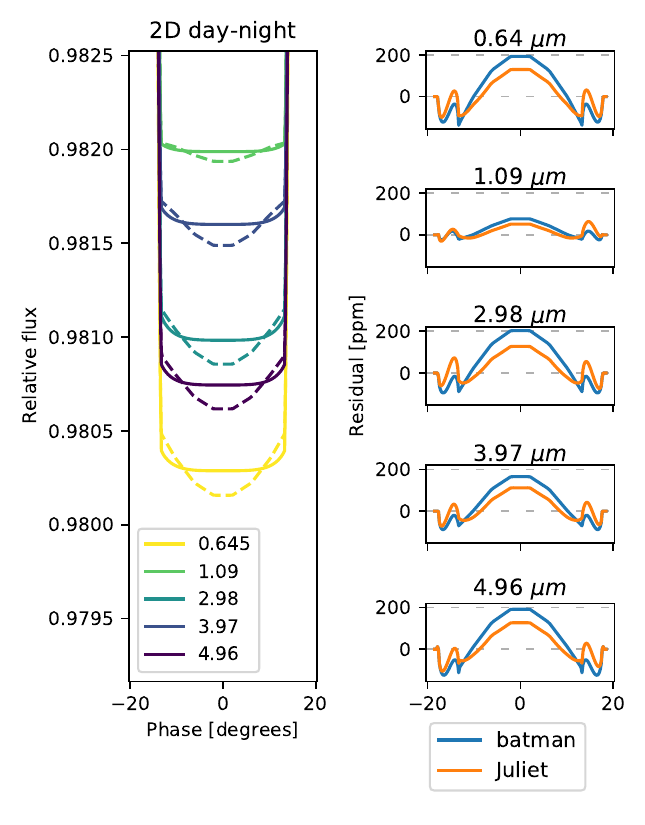}
    \includegraphics[width=.48\textwidth]{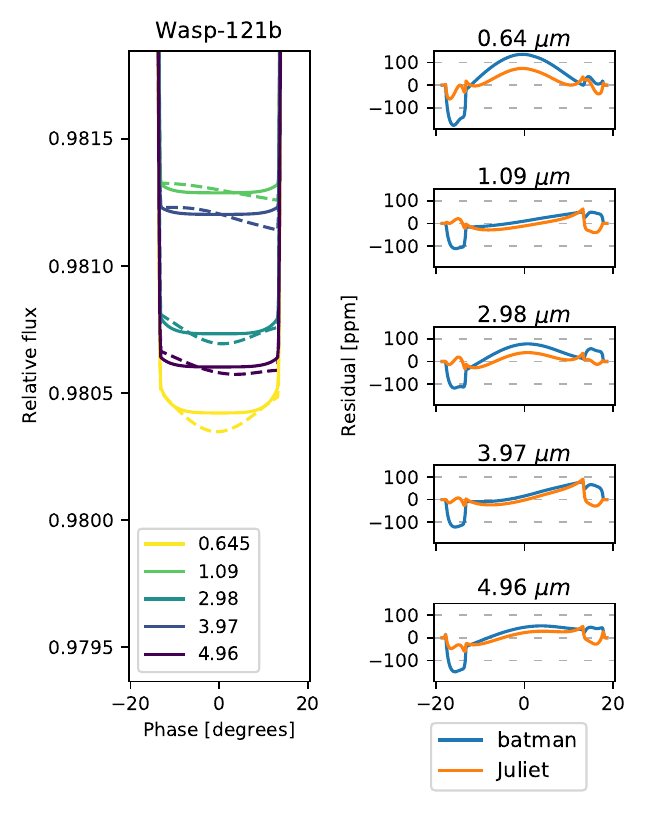}
    \caption{
    For each case (2D day-night and \waspoto:
    Left: Lightcurves generated by \pytmo compared to Juliet (with retrieved limb-darkening, 
    $T_0$ and $R_P$).
    Right: Residuals of Juliet (orange) to \pytmo, with \batman (no limb-darkening, $T_0 = 0$ and $R_P$ taken at mid-transit from \pytmo) given as a reference.
    }
    \label{fig_lc_ld}
\end{figure*}

\fig{fig_lc_ld} shows the differences between the 1D retrieved model, including limb-darkening, and the input lightcurve calculated via \pytmo for \waspoto.
Let us keep in mind here that the ingress/egress reductions are mainly due to the T$_0$ shift, as discussed in \sect{central_time_variations}.
We are therefore more interested in the full-transit.
We can see that the U-shape of the limb-darkening lightcurve (here using a quadratic law), is a better match for most wavelengths, as it can better fit the U-shape or slope of the \pytmo lightcurve.

\begin{figure*} [ht]\centering
    \includegraphics[width=.45\textwidth]{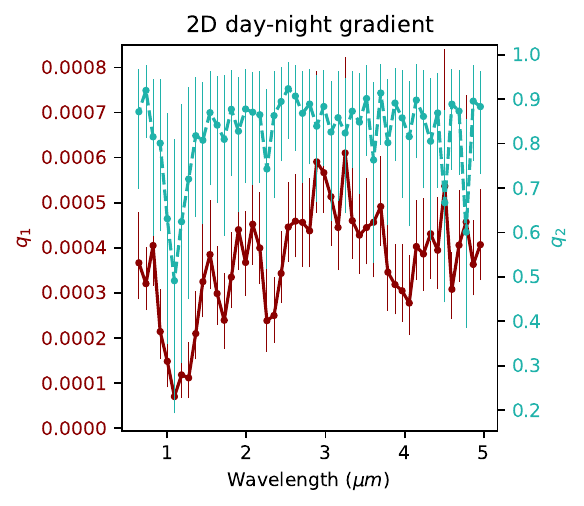}
    \includegraphics[width=.45\textwidth]{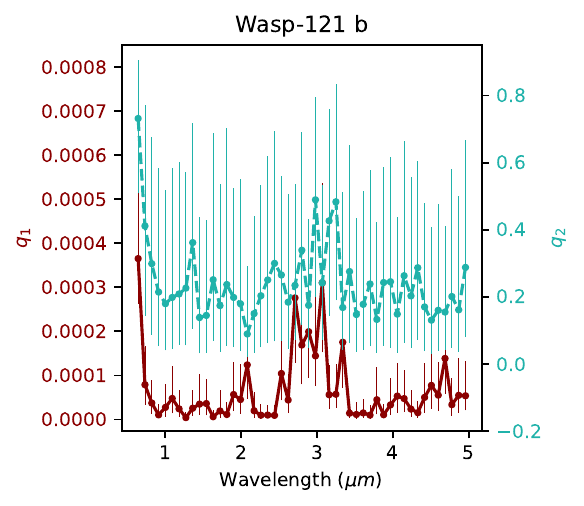}
    \caption{Retrieved values by Juliet for the limb-darkening coefficients $q_1$ and $q_2$ of the 2D day-night gradient case and \waspoto.
    The scale for $q_2$ (indicated on the right) is different in the two plots.
    }
    \label{fig_ld_retrieved_wasp_121_ld}
\end{figure*}

However, this better fit does not represent the original data, in which no limb-darkening has been used.
The limb darkening coefficients retrieved are shown in \fig{fig_ld_retrieved_wasp_121_ld}.
The values of these retrieved coefficients vary from one wavelength to the other.
Around 1~$\mu m$, the shape of the lightcurve is almost flat for the 2D day-night gradient, thus the limb-darkening coefficient is very low.
The shape of the spectral features can be easily distinguished in the values of $q_1$.

The retrieved apparent radius, shown in \fig{fig_rprs_retrieved_wasp_121} is therefore even more biased when it comes to the day-night thermal difference (top plot).
East-west effects of the simulation of \waspoto (bottom plot) seem to reduce this bias.
Indeed, the retrieved values of the limb-darkening coefficients in the case of \waspoto are much smaller than for the 2D day-night gradient, as shown by \fig{fig_ld_retrieved_wasp_121_ld}, and the resulting retrieved lightcurve is consequently closer to a retrieval without limb-darkening.

There are two caveats to this study.
The first one is that we have generated the forward lightcurves without limb-darkening.
This is an implementation issue, as the computation of the intersection area of the transmittance grid with the stellar disc (see \app{sec:intersection}) relies on the assumption that the stellar flux is uniform over the disc.
To take into account the limb-darkening of the stellar flux, one should use a different method, that is, to integrate the flux over each cell by discretizing the cell following the radial dimension, allowing us to plug in the limb darkening law, based on the radial coordinate.
One downside of this integration is its additional cost which would quite slow the overall computation.
This has not been implemented in our model yet.

The second caveat is that these limb-darkening fits have been done using the quadratic law. 
It could be interesting to see how other laws behave, although studying other limb-darkening laws in addition to the generation of limb-darkening forward 3D lightcurves deserves the focus of an entire study and is thus not addressed further here.
We emphasize that the retrieved limb-darkening coefficients depend on the wavelength and we therefore conclude that lightcurves retrievals of \UHJs should be realized over multiple wavelengths to avoid being biased by the day-night gradient of the planet atmosphere.  

\section{Atmosphere vs ellipsoid rotation in the case of \UHJs}
\label{atmosphere_vs_ellipsoid}

\begin{figure}[h]\centering
\includegraphics[width=.5\textwidth]{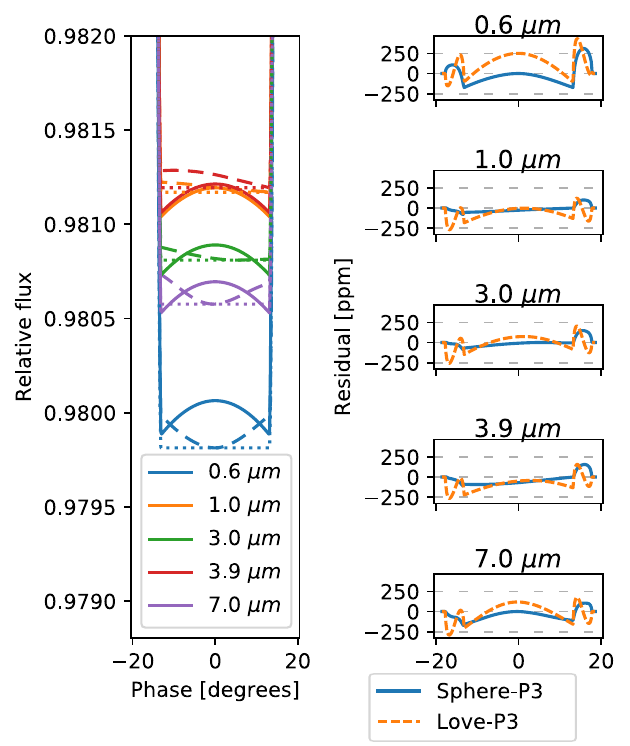}
\caption{Ellipsoid lightcurves compared to lightcurves generated via \pytmo (i.e., spherical core, but atmospheric envelope). Rotation is taken into account for both methods.
Left: Lightcurves generated by \pytmo (dashed) compared to the spherical model (dotted) and the ellipsoid model (solid).
Right: Residuals of the spherical model (Sphere) versus \pytmo (P3), and the ellipsoid model (Love).
}
\label{fig:pytmo_ellc_rotate}
\end{figure}

\cite{Barros2022} proposed to use an ellipsoid model to fit the lightcurve of Wasp-103b, using \verb+ellc+, a tool developed by \cite{maxted2016ellc} that provides an ellipsoid lightcurve model which also offers the ability to do a Bayesian analysis.
They suggested the model fitted the data better than a spherical model.
However, as we can see in \fig{fig:pytmo_ellc_rotate}, 
the lightcurve generated by \pytmo (to be noted, with a spherical core) is better fitted by the spherical model.
We used a Love number equal to 1.39 for this study-case corresponding to \waspoto \citep{Hellard_2020}.

\begin{figure*}[ht]\centering
\hspace*{1.5cm}
\includegraphics[height=25mm]{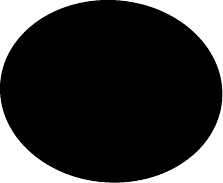}
\hfill
\includegraphics[height=25mm]{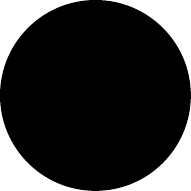}
\hfill
\includegraphics[height=25mm]{img/ellipse.pdf}
\hspace*{1.5cm}
\caption{Projection of an ellipsoid for early, mid and late transit phases (not to scale).
To be compared to \fig{fig_transmittances}.
}
\label{fig_ellipses_transit}
\end{figure*}

The deformation of the ellipsoid has the reverse shape than that of the atmospheric signal of the day-night temperature gradient (see lightcurve at 0.6~$\mu m$ for example).
This is due to the apparent surface projected onto the plane of the sky.
As shown in \fig{fig_ellipses_transit}, the ellipsoid covers more surface during the early and late stages, compared to the mid-transit stage.
The atmospheric signal due to the day-night dichotomy of \UHJs has a reverse effect, as mentioned before: the atmosphere covers more surface during the mid-transit stage due to the hot-day side being visible all around the limb, while the night-side is more visible during the early and late stages, decreasing the scale-height of the atmosphere and the apparent surface (see \fig{fig_transmittances}).

For \HJs, when the atmospheric signal is more influenced by east-west effects rather than day-night thermal differences, this is less visible.

We can therefore only conclude that if the ellipsoid model is indeed fitting better the original data of Wasp-103b, the actual Love number found may be underestimated due to the signal linked to the atmosphere, and more especially if there is a strong day-night dichotomy.
A coupling between a tidally-deformed core and an atmospheric model would be needed to study both effects simultaneously.

\section{Conclusion}

In this paper, we have introduced a novel framework to compute lightcurves from a 3D GCM simulation.
The parameters for the model include the planet's orbit radius, period (optional, but necessary for the conversion between phase and time) and inclination, as well as the number of phases and transmittances for which to compute the lightcurve.
The documentation for this tool, \href{https://forge.oasu.u-bordeaux.fr/jleconte/pytmosph3r-public}{\pytmo~2.2}, is available \href{http://perso.astrophy.u-bordeaux.fr/~jleconte/pytmosph3r-doc/index.html}{here}\footnote{\url{http://perso.astrophy.u-bordeaux.fr/~jleconte/pytmosph3r-doc/index.html}
\label{pytmodoc}
}.

We have used \pytmo to generate lightcurves in either 2D (with a day-night temperature gradient) representative of an \UHJ, or in 3D (with simulations of \waspoto and \wasptn) and studied how 1D models behave in retrieving such data.
The 1D lightcurve retrieval model used is Juliet \citep{espinoza2019juliet}, which is based on the 1D lightcurve model \batman \citep{kreidberg2015batman}. 

Our conclusions are that the 2D day-night dichotomy of an \UHJ has two impacts on the lightcurve: 1) the planet apparent radius decreases in the early and late stages of the transit, compared to the mid-transit, leading to a U-shape lightcurve, and 2) the planet apparent radius increases during the ingress/egress due to the day-side obscuring the star longer than the night-side.

The lightcurve's U-shape due to the day-night thermal dichotomy may be confused with limb-darkening.
However, the limb-darkening coefficients depend on the wavelength, and could therefore be distinguished from the atmospheric signal in this way.
East-west asymmetries might also allow to break this degeneracy.

Furthermore, we find that the atmospheric signal of a day-night strong temperature gradient has a signature opposite to the tidal deformation of the planet core (not taking into account east-west asymmetries) and therefore the Love number retrieved by an ellipsoid model may be underestimated if not taking into account the atmospheric signature in the lighcturve.

The east-west asymmetries of the \UHJ simulation (\waspoto) are also visible in the lightcurve data, and will mostly translate as a shift in the time of mid-transit (or central transit time) T$_0$ for the lightcurve retrieval code.
The hostpot being shifted to the east in our simulation, the planet apparent radius appears larger during the late stage/egress compared to the early/ingress stages.
We have included TiO/VO in our simulations.
However, recent studies show that \waspoto might be depleted in TiO/VO \cite{hoeijmakers2022mantis}.
This does not change the general conclusions of the study.

The east-west asymmetries are stronger in the case of the \HJ \wasptn, leading to a larger shift in the central transit time T$_0$ retrieved by the 1D retrieval model.
This simulation of \wasptn is linked to a shift that can be as large as 20~seconds, and which varies from one wavelength to the other.
Between 1 to 5 $\mu m$, the amplitude of these variations is around 17~seconds.
This is less than the real observations by the JWST \citep{rustamkulov2023early}, which could be due to the chemistry or clouds structure in the atmosphere.

A very promising tool that could better describe the spatial deformations of the atmosphere comes under the form of transmission strings, namely \verb+harmonica+ \citep{grant2022transmission}.
In this tool, the planet radius is described as a function of the angle around the limb, parameterized in terms of Fourier series. 
Further studies could investigate how well this model may retrieve a full 3D GCM lightcurve generated via \pytmo.

\begin{acknowledgements}
This work has received funding from the European Research Council (ERC) under the European Union's Horizon 2020 research and innovation programme (grant agreement n$^\circ$679030/WHIPLASH).
AF would also like to acknowledge financial support by LabEx UnivEarthS (ANR-10-LABX-0023 and ANR-18-IDEX-0001) and by the CNES (Centre National d'Études Spatiales).
\end{acknowledgements}

\bibliographystyle{aa}
\bibliography{biblio}

\appendix

\section{Area of intersection between transmittance cells and the stellar disc}
\label{sec:intersection}

\begin{figure}[ht]\centering
\includegraphics[width=.38\textwidth]{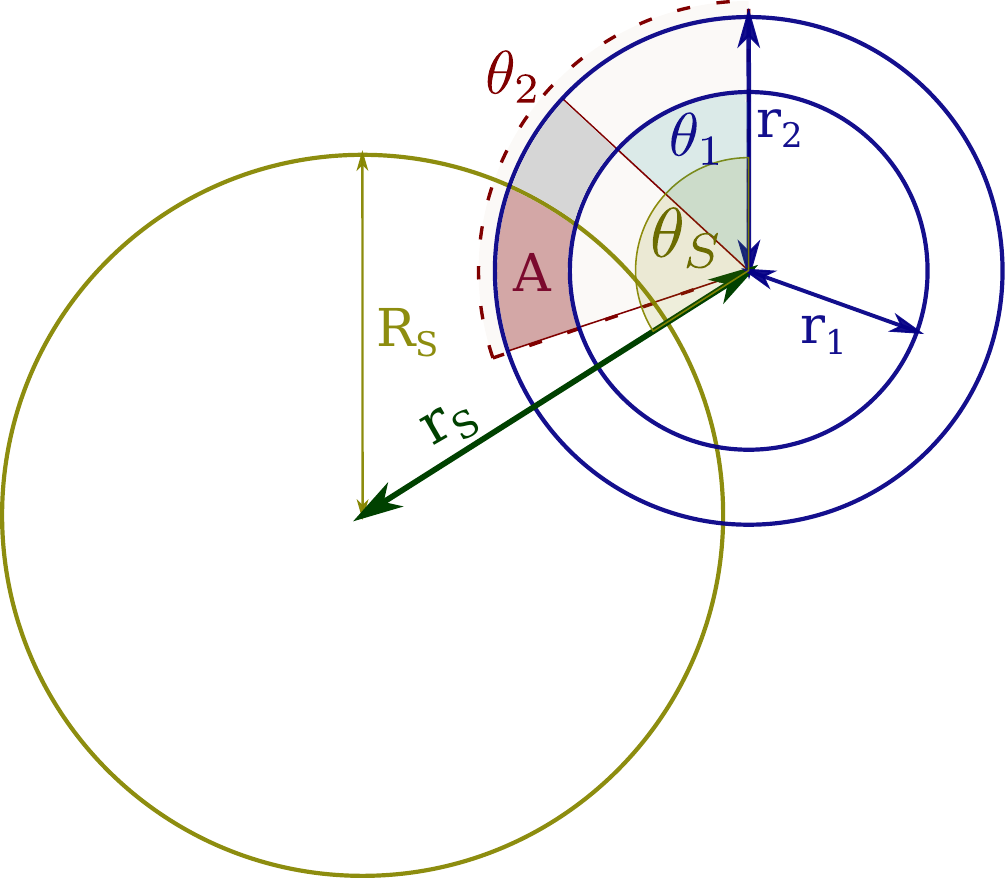}
\caption{Area of intersection $A$ (in red) of one cell $(\ra, \rb, \aa, \ab)$ of the transmittance grid with a stellar disc of radius $Rs$. The star center is located at $(\Ds, \As)$. The coordinates follow the transmittance coordinate system $\coordtrans$.}
\label{fig:exiting_limb_geometry_star}
\end{figure}

The transmittance map has been defined using the polar coordinate system $\coordtrans$, and using a grid of $\nradial$ radial points and $\ntheta$ angular points.
To properly evaluate the opacity of each cell in this grid, defined by its boundaries $\cell$, we need to calculate the surface of intersections between the star apparent disc and this cell $\Sp$ (see \eq{eq:integral}).
This is illustrated by \fig{fig:exiting_limb_geometry_star}.

\begin{figure}[ht]\centering
\includegraphics[width=.27\textwidth]{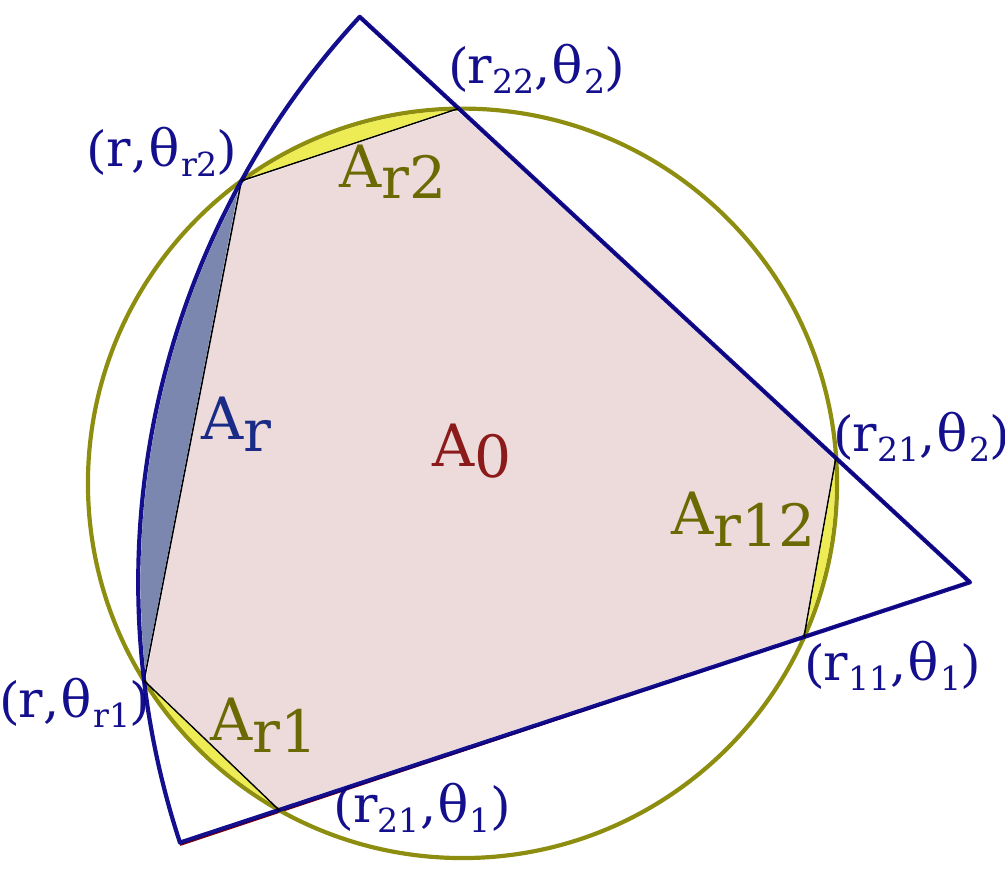}
\caption{Intersection of one sector $(\rc, \aa, \ab)$ of the transmittance grid (boundaries in blue) with the star (boundaries in yellow).
The intersection area can be subdivided into the sum of circular segments (yellow when the circle is the star, blue when the circle is the planet) and triangles (subparts of $A_0$, in red).
Here, the maximum number of intersections have been displayed.
\fig{fig:all_geometry_sector} shows examples of cases with less intersections.
}
\label{fig:exiting_limb_geometry_sector}
\end{figure}

\begin{figure}[ht]\centering
\includegraphics[width=.24\textwidth]{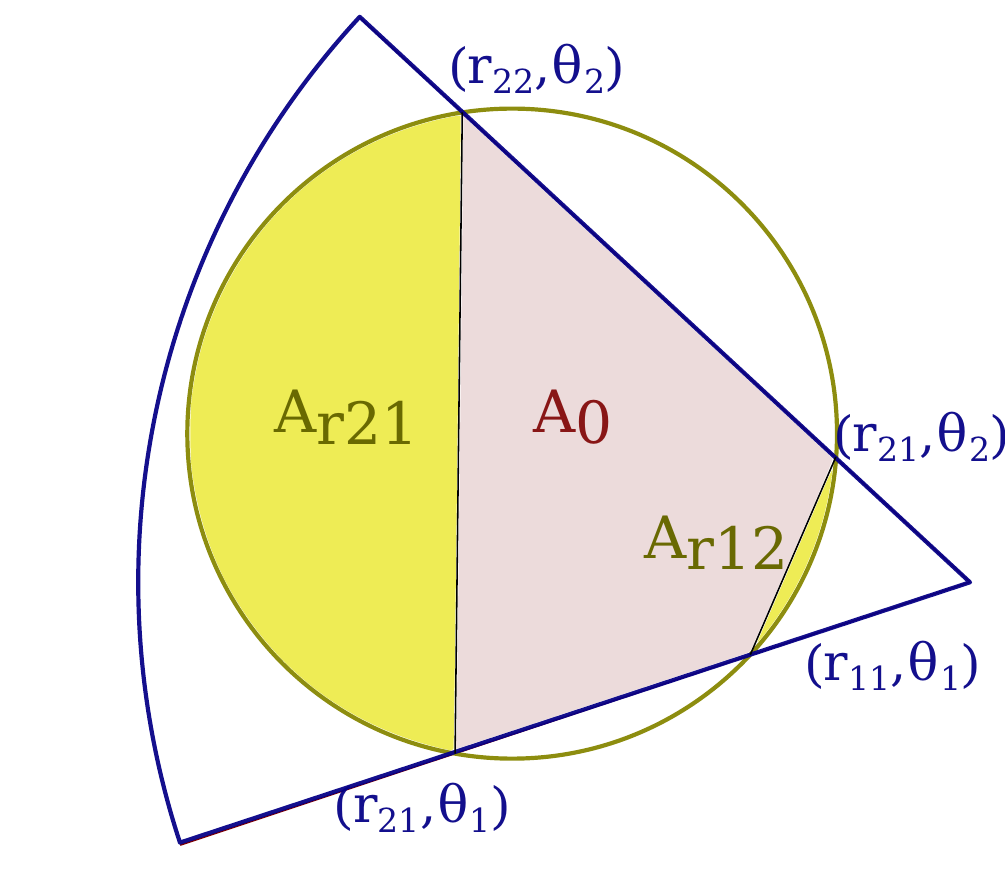}
\includegraphics[width=.24\textwidth]{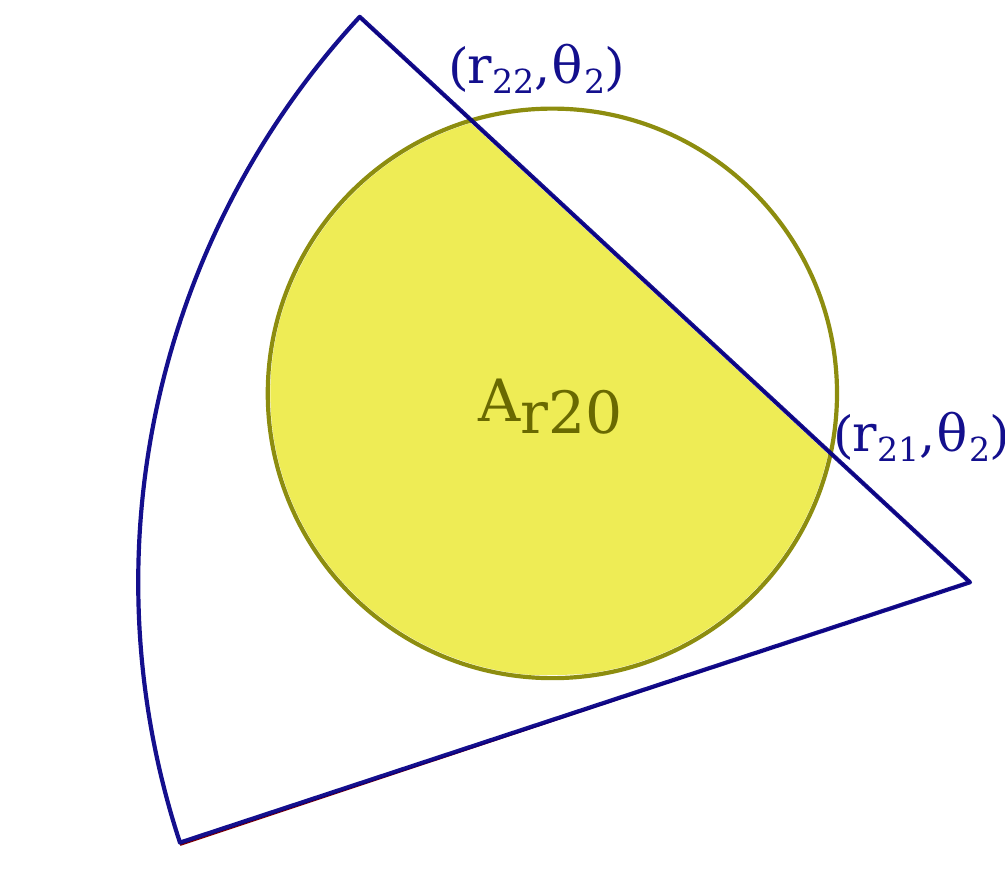}
\includegraphics[width=.24\textwidth]{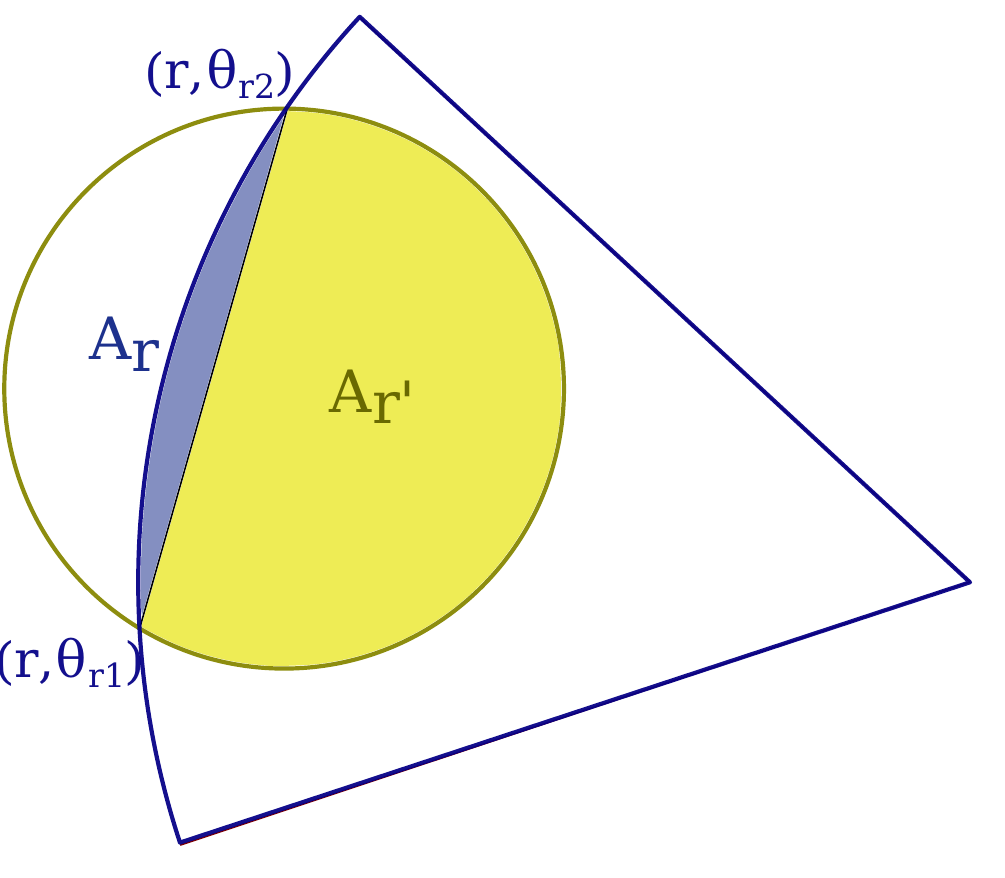}
\includegraphics[width=.24\textwidth]{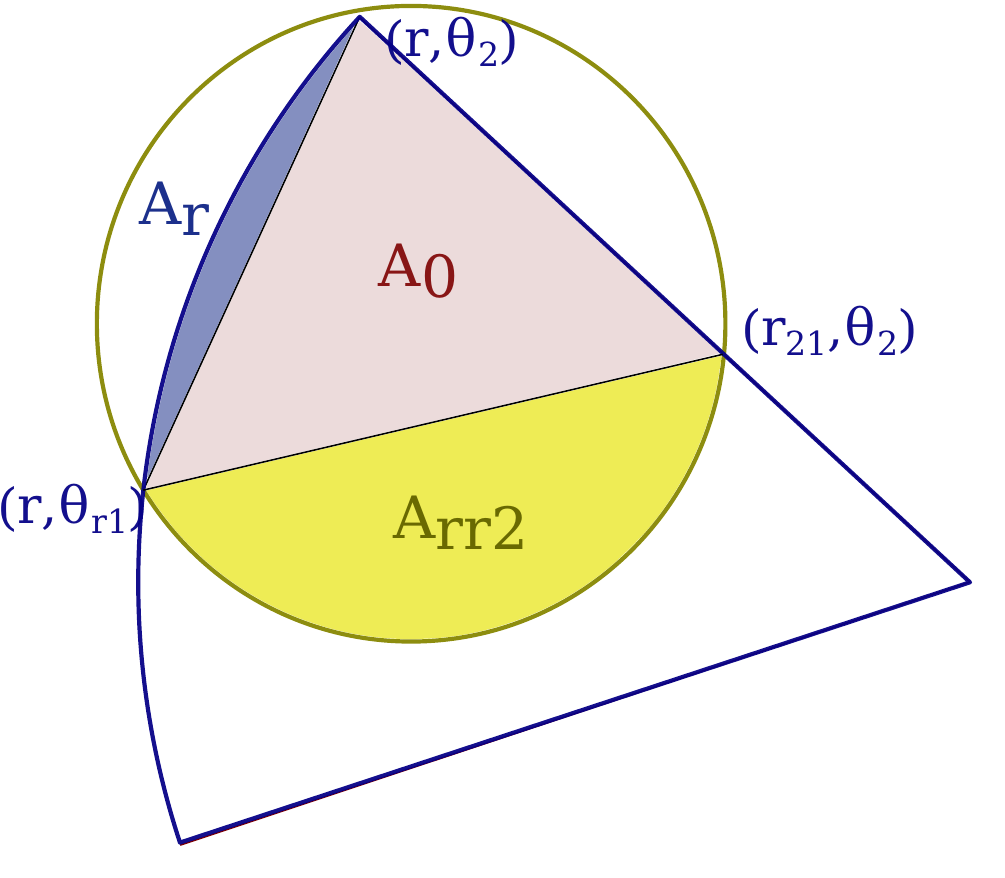}
\caption{Examples of cases with fewer intersections between the sector $(\rc, \aa, \ab)$ and the star. Some areas have merged together.
}
\label{fig:all_geometry_sector}
\end{figure}

The star projected center is defined as the point at coordinates $(\Ds, \As)$ in $\coordtrans$.
Its Cartesian coordinates are $x_S = D \sin(\As)$ and $y_S = D \cos(\As)$.
The boundary of the stellar disc is defined as a circle of radius $\Rs$ around that point.
The points $(0,0)$, $(\Ds,As)$ and $(\rc,\theta)$ form a triangle whose sides are equal to $\Ds$, $\rc$ and $\Rs$.
The law of cosines states:
\begin{equation}
  {\Rs}^2 - {\rc}^2 - {\Ds}^2 + 2\rc\cdot\Ds\cos(\theta - \As) = 0.
  \label{eq:law_cosines}
\end{equation}
For a fixed angle $\theta$, we can rewrite \eq{eq:law_cosines} as:
\begin{equation}
  {\rc}^2 + b \cdot \rc + c = 0,
  \label{eq:find_r}
\end{equation}
with $b = -2\rc \cdot \Ds\cos(\theta - \As) $ and ${c = {\Ds}^2-{\Rs}^2}$,
which is a polynomial equation of degree 2 which can easily be solved to get $\rc$.
The angle $\theta$ can be computed from a fixed $\rc$ by rewriting \eq{eq:law_cosines} as:
\begin{equation}
  \theta = \As \pm \arccos\left(\frac{{\Rs}^2 - {\rc}^2 - {\Ds}^2}{-2\rc\cdot\Ds}\right).
  \label{eq:find_theta}
\end{equation}
Note that for both \eqs{eq:find_r}{eq:find_theta}, there are zero, one or two solutions.
We apply \eqs{eq:find_r}{eq:find_theta} to find all intersections of the star boundary (circle of radius \Rs) with the radial and angular points of the transmittance grid.
A cell of the transmittance grid is defined using its boundaries $\cell$.
The area of the cell $\cell$ can be expressed as:
\balign{
  &A_\textnormal{cell}(\ra, \rb, \aa, \ab) =  \nonumber\\
  &|A_\textnormal{sector}(\rb, \aa, \ab)- A_\textnormal{sector}(\ra, \aa, \ab)|.
  \label{eq:cell_from_sector}
}
See \fig{fig:exiting_limb_geometry_star} for an illustration of the intersection area.
Therefore, following \eq{eq:cell_from_sector}, we now reduce the problem to finding the area of a sector of the transmittance grid $A_\textnormal{sector}(\rc, \aa, \ab)$, as illustrated in \fig{fig:exiting_limb_geometry_sector}.
As shown in this figure, this intersection can be expressed as the sum of simpler geometric objects, composed of circular segments and triangles.
We express below the intersection for the maximum number of intersections.
If an intersection is outside the sector, we can simply select the nearest corner instead.
If there are no intersections (no solution to either \eq{eq:find_r} or \eq{eq:find_theta}), two circular segments will join.
Examples of such cases are shown in \fig{fig:all_geometry_sector}.

The area of a triangle can be computed from its sides $a$, $b$ and $c$ using Heron's formula:
\begin{equation}
  A_\textnormal{triangle}(a,b,c) = \sqrt{s(s-a)(s-b)(s-c)},
  \label{area_triangle}
\end{equation}
where $s = (a+b+c)/2$.

The area of a circular segment of angular width $\alpha$ for a circle of radius $R$ is given by:
\begin{equation}
  A_\textnormal{segment}(\alpha,R) = \left|\frac{\alpha - \sin(\alpha)}{2} R^2\right|,
  \label{area_segment}
\end{equation}

Finally, the core of the planet is considered completely opaque, and therefore, the surface area of intersection of a core of radius $\Rp$ with the stellar disc of radius $\Rs$, of which the centers are at a distance $\Ds$, is equal to the area of intersection of two discs.
If $d \geq \Rs + \Rp$, then the surface area of intersection is zero.
Else, if $d \geq |\Rs - \Rp|$ the surface is equal to the area of the smallest circle.
Else, we may use:

\balign{
  A_\textnormal{circles}(\Rp, \Rs, d) = & \Rp^2 \arccos \left( \frac{A_P}{\Rp^2} \right) \nonumber\\
  & + \Rs^2 \arccos \left( \frac{A_S}{\Rs^2} \right) \nonumber\\
  & - A_P \sqrt{ \Rp - {A_P}^2  }  \nonumber\\
  & - A_S \sqrt{ \Rs - {A_S}^2  }
    \label{eq:intersection_circles}
}
where
\begin{eqnarray}
A_P=\frac{\Rp^2-\Rs^2+d^2}{2d},\nonumber\\
A_S=d - A_P.
\end{eqnarray}

These equations are equivalent to Eq.~(3) from \cite{kreidberg2015batman}.

\section{Convergence \& computational cost}
\label{sec:convergence}

We have previously validated the computation of transmission spectra in \pytmo in \cite{falco2022toward} by studying the stability of the model when changing the position of the observer in the frame of reference of the planet.

We have also validated our lightcurve implementation using \batman, in the case where no atmosphere is present, but this comes down to validating the equations calculating the intersection area of two circles, i.e., \eq{eq:intersection_circles}, or its equivalent, Eq.~(3) in \cite{kreidberg2015batman}.

We have in addition studied the convergence of the lightcurve module in \pytmo.
Studying a case symmetric along the star-observer axis (translation 2D day-night temperature map defined in \eq{eq:temperature2D}, or 1D), the number of angular points in the transmittance map ($\ntheta$) should have no effect on the resulting integral.
We have observed an accuracy of less than 1~ppm for all $\ntheta$ for the 2D case of \tab{tab:2D_case}.
The parameters used for generating lightcurves with \pytmo are listed in \tab{tab_lc_params}.
\begin{table}[ht]\centering
  \begin{tabular}{|cccc|}\hline
  $\Np$ & $\Negress$ & $\Ningress$  & Orbit radius \\\hline
  100 & 30 & 30 & 0.0254 au\\
  \hline
  \end{tabular}
  \caption{\pytmo parameters used for generating lightcurves.}
  \label{tab_lc_params}
\end{table}

\begin{figure}[ht]\centering
\includegraphics[width=.5\textwidth]{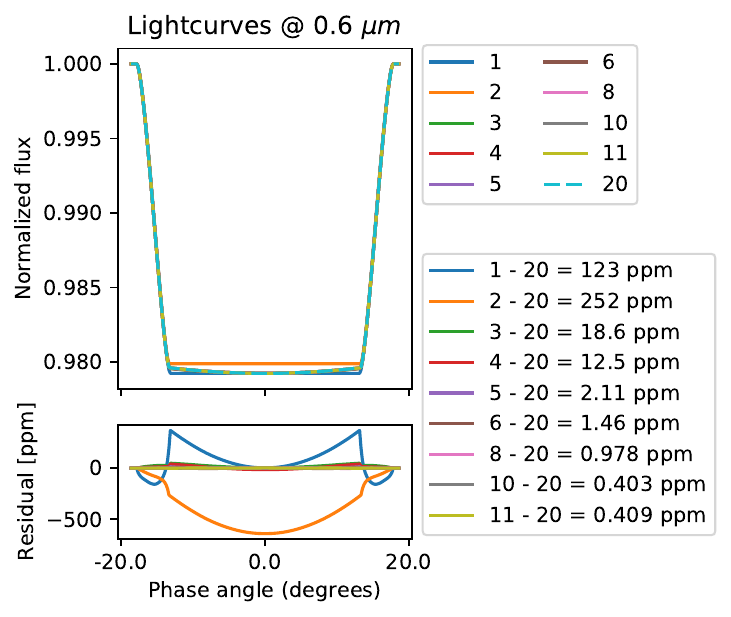}
\includegraphics[width=.5\textwidth]{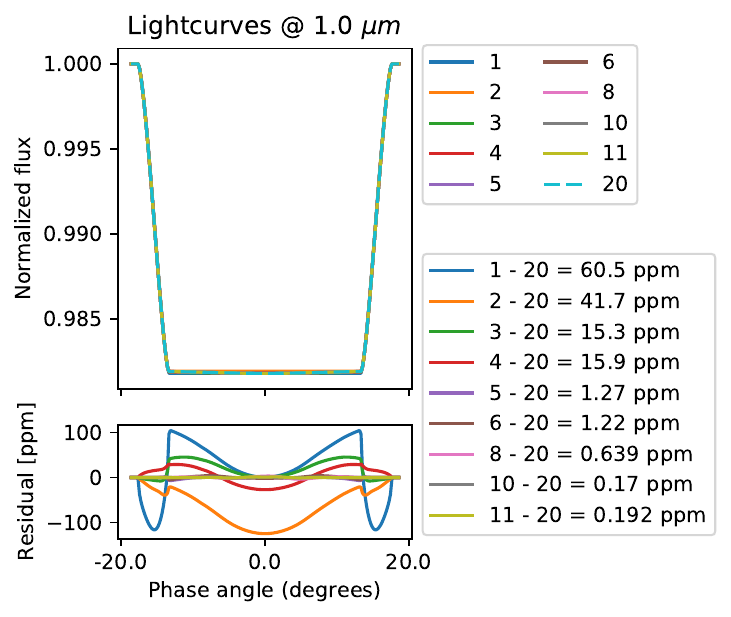}
\caption{Increasing the number of steps for which we compute the transmittance increases the accuracy of the computed lightcurve.
The legend indicates the number of transmittance maps $\Nt$ used, as well as their difference to $\Nt = 20$ and the corresponding standard deviation.
The standard deviation for $\Nt = 8$ is less than 1~ppm and is therefore an acceptable number compared to current instrumental errors \citep{rustamkulov2023early}.
The 2D study-case has been used for this experiment (\tab{tab:2D_case}).
}
\label{fig:lc_2D_convergence}
\end{figure}

We have also introduced a number of transmittance maps $\Nt$ to calculate all along the transit in \sect{sec:transit_model}.
For the rest of the $\Np$ phases, the transmittance map is interpolated from the $\Nt$ maps.
We assume the \Nt maps to be regularly distributed in the phase/time space.
This induces an error in the resulting lightcurve.
\fig{fig:lc_2D_convergence} shows the errors one can expect when using a number of transmittance maps $\Nt$ too low for a \HJ (2D study-case of \tab{tab:2D_case}).
With $\Nt = 1$, we can actually see the effect of \emph{not} including rotation in the model, which is not negligible.
This is discussed more at length in \sect{sec:rotation}.

\begin{figure}[ht]\centering
\includegraphics[width=.48\textwidth]{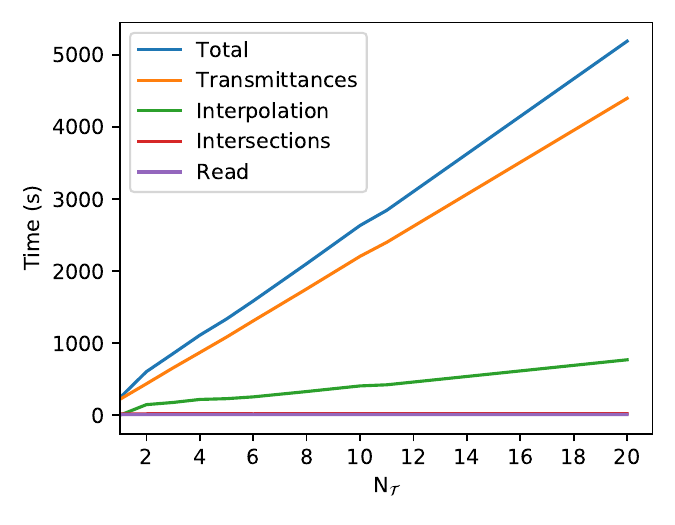}
\caption{Cost of increasing the number of transmittances $\Nt$ computed for 30000 wavelengths.
The 2D study-case has been used for this experiment (\tab{tab:2D_case}).
}
\label{fig:times_2D_convergence}
\end{figure}

Computing one transmittance map scales with the number of molecules, the spectral resolution and the atmospheric resolution used. 
We can see in \fig{fig:times_2D_convergence} how interpolating between the $\Nt$ transmittances can prove to be very effective.

Overall, $\Nt = 8$ seems to be an acceptable trade-off between computational time and accuracy.
It seems even acceptable to use $\Nt = 5$ though this might prove to be inadequate in some cases.

\section{Time/phase resolution in lightcurve retrieval}
\label{phase_resolution}

When generating a lightcurve with \pytmo, the default option will discretize the ingress/egress with a higher resolution (in the time, or phase, dimension), compared to the rest of the transit, as a trade-off between accuracy and computational time.
We therefore discuss here the difference between lightcurves:
\begin{enumerate}
    \item 
    generated using \pytmo with no resampling, here with parameters from \tab{tab_lc_params}, i.e, $\Negress = 30$ and $\Np = 100$, and
    \item
    interpolated from the above over $\Np = 300$ phases equally distributed. 
    \end{enumerate}

\begin{figure}[ht]\centering
    \includegraphics[width=.5\textwidth]{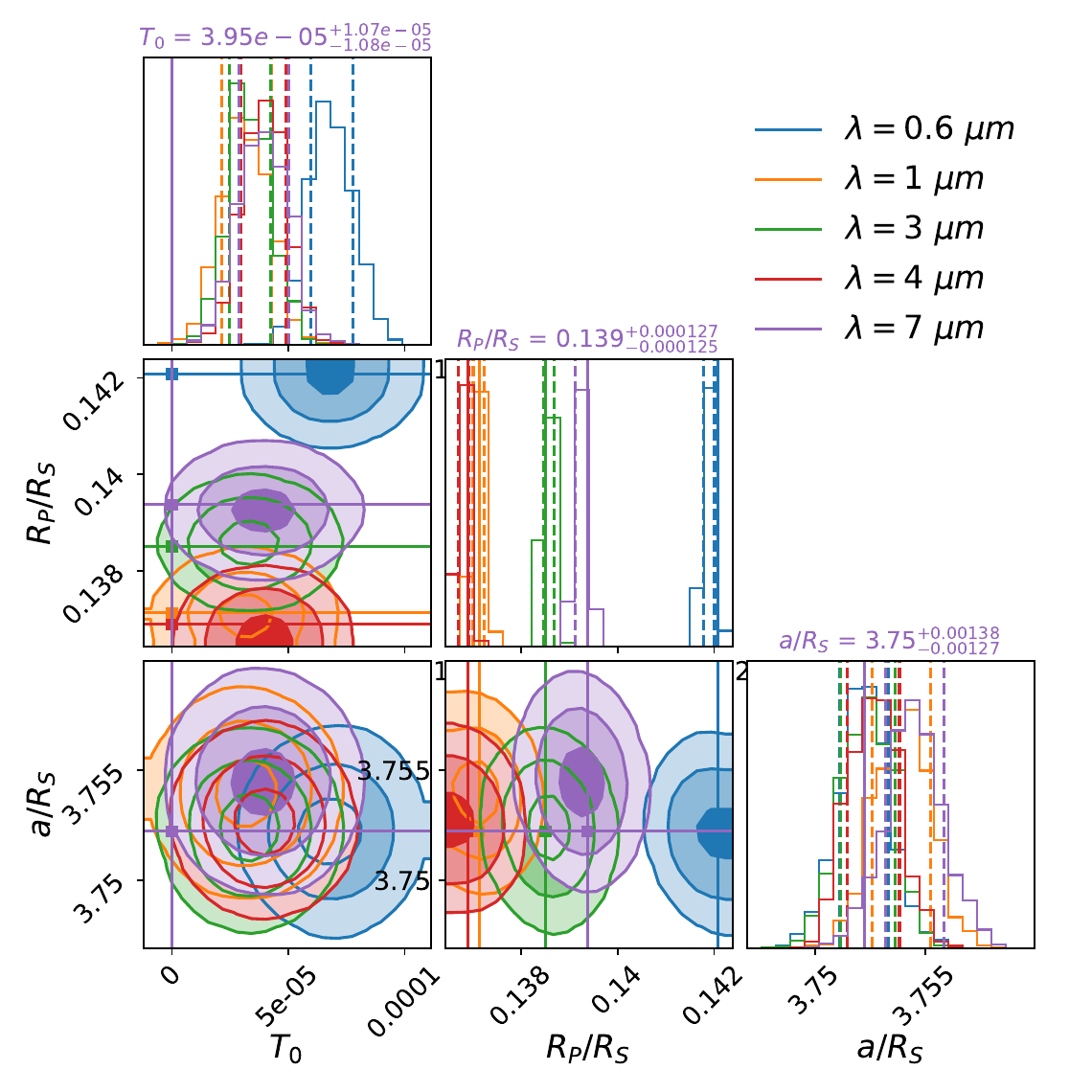}
    \includegraphics[width=.5\textwidth]{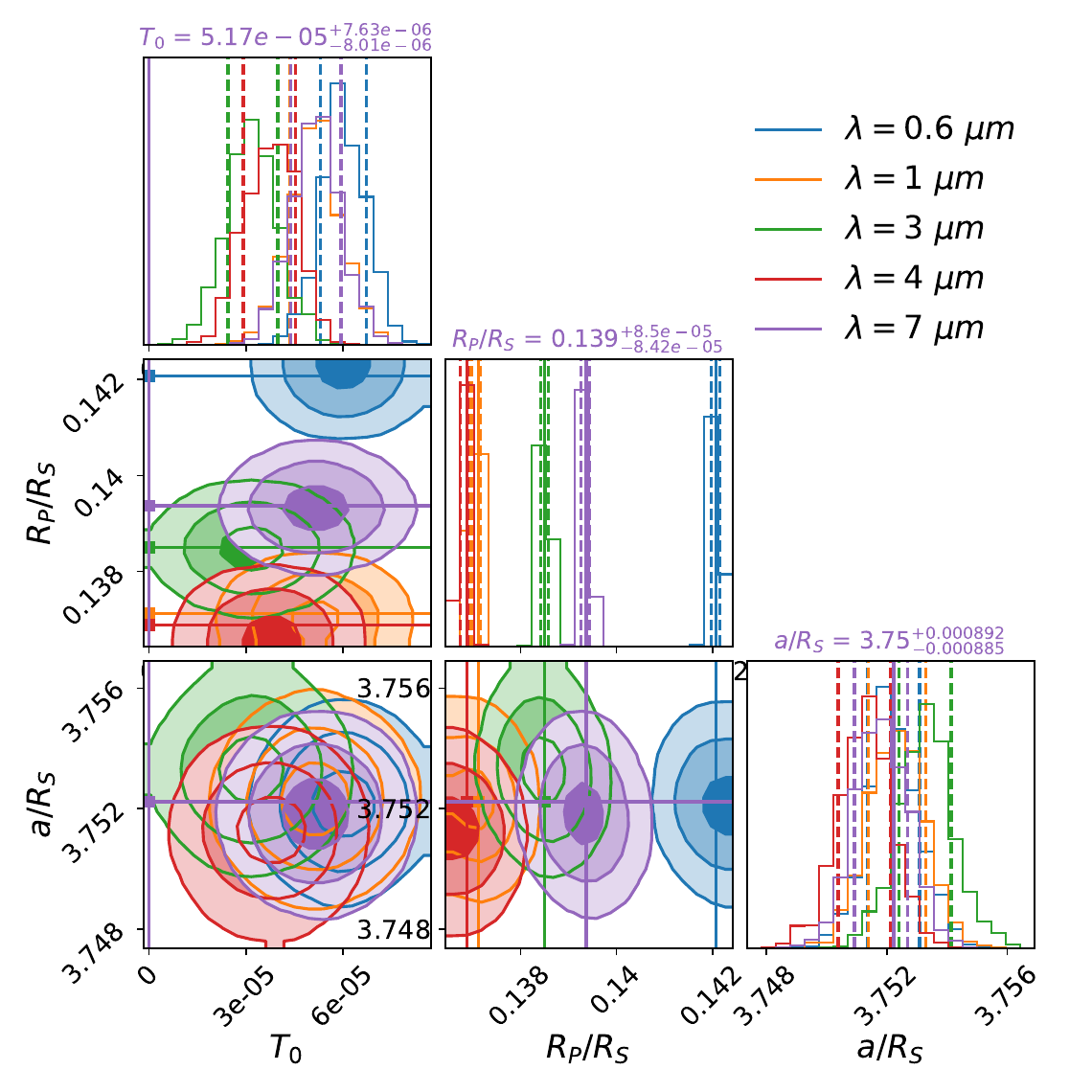}
    \caption{Values retrieved by Juliet for the lightcurve generated by \pytmo in the 3D case (\fig{fig:wasp}), without rotation, at wavelengths 0.6, 1, 3, 3.9 and 7 $\mu m$.
    The parameters for the retrieval are described in \tab{tab:parameters_juliet}.
    Parameters for the input lightcurve (\pytmo) are given in \tab{tab_lc_params}.
    Top: $\Negress = 30$ and $\Np = 100$.
    Bottom: Interpolation over $\Np = 300$ phases.
    }
    \label{fig:corner_juliet_fit_norebin}
\end{figure}

\begin{figure*}[ht]\centering
    \begin{subfigure}{.49\textwidth}
    \includegraphics[width=\textwidth]{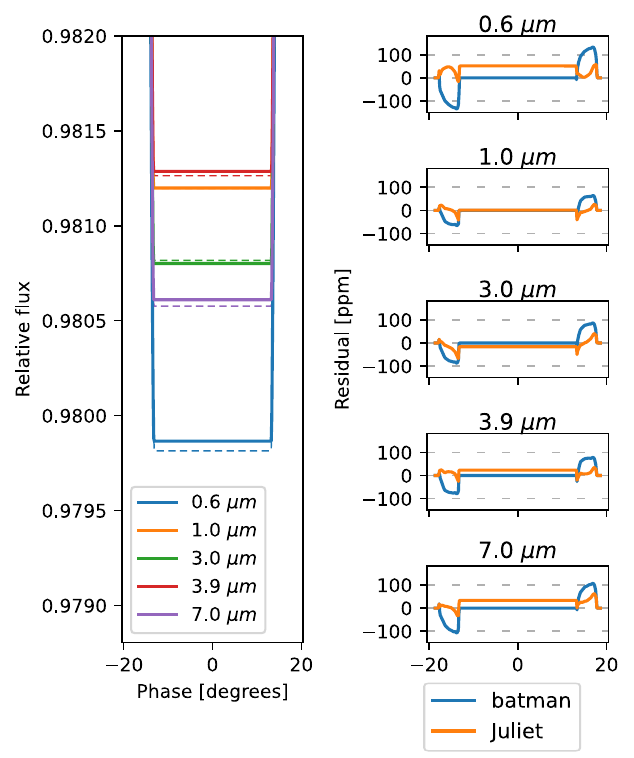}
    \caption{$\Negress = 30$ and $\Np = 100$}
    \label{juliet_norebin}
    \end{subfigure}
    \begin{subfigure}{.49\textwidth}
    \includegraphics[width=\textwidth]{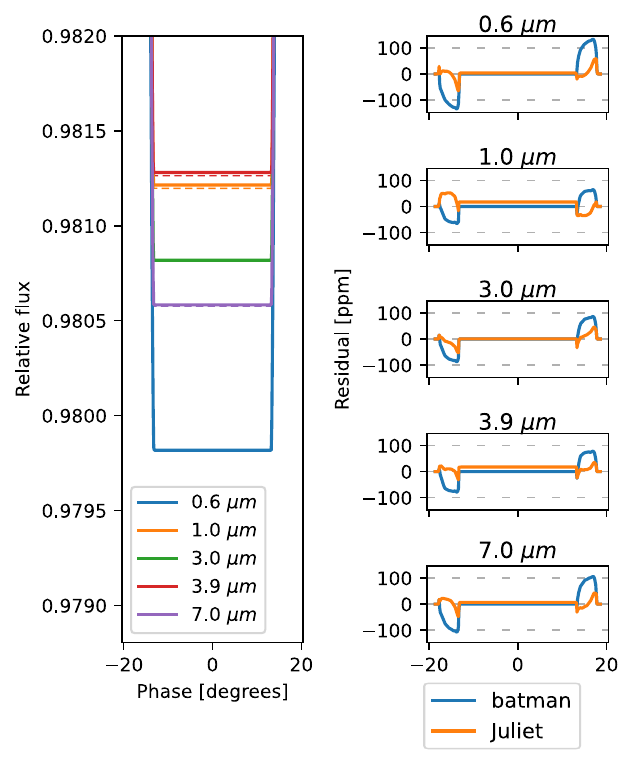}
    \caption{Interpolation over $\Np = 300$ phases}
    \label{juliet_interpolation}
    \end{subfigure}
    \caption{Comparison between \pytmo, \batman ($T_0 = 0$) and Juliet (values retrieved in \fig{fig:corner_juliet_fit_norebin}).
    For each subfigure:
    Left: Lightcurves generated by \pytmo (dashed) compared to Juliet (solid).
    Right: Residuals of \batman (blue) and Juliet (orange) to \pytmo.
    The residual of Juliet is lower in average than that of \batman, mainly because of the shift in T$_0$.
    Using $\Negress = 30$ and $\Np = 100$ leads to a statistical over-representation of the egress/ingress in the retrieval, which can be corrected using an interpolation over $\Np = 300$ phases (see \fig{juliet_fit_norebin}).
    }
    \label{juliet_fit_norebin}
\end{figure*}

The values retrieved by Juliet are shown in \fig{fig:corner_juliet_fit_norebin} for both cases, and the corresponding lightcurves are shown in \fig{juliet_fit_norebin}.

The egress/ingress (which last a shorter amount of time that the rest of the transit) is over-represented from a statistical point of view in case 1).
This leads to a bias in the retrieval, in which the algorithm will try to minimize the egress/ingress more that the rest of the transit.
To avoid this statistical bias, we can interpolate the signal over a finer phase grid, or we can change the weight of each point using the error-bars used in the retrieval.
We have used the first method here, though they are equivalent from a statistical point of view.

We can see that in case 1) in \fig{juliet_norebin}, the apparent radius during the main transit (when the planet, including its atmosphere, is fully in front of the star, excluding the ingress/egress) is underestimated by Juliet by more than 50~ppm at 0.6~$\mu m$.
This is more than the maximum residual during ingress/egress, though this varies from wavelength to wavelength.

In case 2), i.e., \fig{juliet_interpolation}, the maximum residual always occurs during ingress/egress, and more precisely when only a small portion is either in front or out of the star (due to the atmosphere opacity, in a similar way as explained in \fig{fig:batman_vs_pytmo} and \fig{fig:diffuse_spherical_opacity}).

In conclusion, we highly recommend a user of \pytmo to be aware of this statistical bias.
When computing a lightcurve with a higher resolution over the ingress/egress (for computational reasons), one should always either interpolate it over an equally-distributed sample, or change the error-bars accordingly, before using it in retrieval methods.

This study does not take into account other biases that unfold when changing the temporal binning of lightcurves and we refer interested readers to \cite{kipping2010binning} for more information.

\end{document}